\documentclass[useAMS,usenatbib]{mnras}

\usepackage[utf8]{inputenc}
\usepackage{graphicx}
\usepackage{color}
\usepackage{booktabs}
\usepackage[nolist]{acronym}
\usepackage{amssymb}
\usepackage[]{natbib}
 
\usepackage{array}
\usepackage{multirow}
\usepackage{hhline,amsmath}
\usepackage{makecell}

\newcommand{\eagle}{{\sc eagle}}
\newcommand{\gadget}{{\sc gadget}}

\def\apj{ApJ}
\def\apjs{ApJS}
\def\apjl{ApJ}

\def\mnras{MNRAS}

\def\nat{Nature}

\def\aap{A\&A}
\def\pasp{PASP}

\title[The impact of assembly bias on the halo occupation]{The impact of assembly bias on the halo occupation in hydrodynamical simulations}
\author[Artale, M. C.]{M. Celeste Artale$^{1,2}$\thanks{E-mail:Maria.Artale@uibk.ac.at, mcartale@gmail.com}, Idit Zehavi$^{3}$, Sergio Contreras$^{4,5}$ \& Peder Norberg$^{6}$ 
\\
$^{1}$Institut f\"{u}r Astro- und Teilchenphysik, Universit\"{a}t Innsbruck, Technikerstrasse 25/8, 6020 Innsbruck, Austria\\ 
$^{2}$CONICET-Universidad de Buenos Aires. Instituto de Astronom\'ia y F\'isica del Espacio (IAFE), CC 67, Suc. 28, 1428 Buenos Aires, Argentina\\
$^{3}$Department of Physics, Case Western Reserve University, 10900 Euclid Avenue, Cleveland, OH 44106, USA\\
$^{4}$Instituto Astrof\'isica, Pontifica Universidad Cat\'olica de Chile, Santiago, Chile\\
$^{5}$Centro de Estudios de F\'isica del Cosmos de Arag\'on (CEFCA), Plaza San Juan 1, Planta-2, Teruel, E-44001, Spain\\
$^{6}$ICC \& CEA, Department of Physics, Durham University, South Road, Durham, DH1 3LE, UK\\
} 
 
\begin{document}
\maketitle

\begin{abstract}
We investigate the variations in galaxy occupancy of the dark matter haloes 
with the large-scale environment and halo formation time, using  
two state-of-the-art hydrodynamical cosmological simulations, \eagle\ and Illustris. 
For both simulations, we use three galaxy samples with a fixed number density ranked by stellar mass.
For these samples we find that low-mass haloes in the most dense environments are more likely to host
a central galaxy than those in the least dense environments.
When splitting the halo population by formation time, these relations are stronger. Hence, at a fixed low halo mass,
early-formed haloes are more likely to host a central galaxy than late-formed haloes since they have had more time 
to assemble. The satellite occupation shows a reverse trend where early-formed haloes host
fewer satellites due to having more time to merge with the central galaxy. 
We also analyse the stellar mass -- halo mass relation for central galaxies 
in terms of the large-scale environment and formation time of the haloes.
We find that low mass haloes in the most dense environment host relatively more massive central galaxies.
This trend is also found when splitting the halo population by age, with early-formed haloes hosting more massive galaxies. 
Our results are in agreement with previous findings from semi-analytical models, providing robust predictions for the 
occupancy variation signature in the halo occupation distribution of galaxy formation models. 

\end{abstract}



\section{Introduction} \label{sec:intro}

In the current concordance $\Lambda$CDM paradigm, galaxies form and evolve in dark matter haloes.
Studying the relation between galaxies and haloes is crucial for a better understanding of
galaxy formation and for constraining cosmological models.
Different methods have been developed to investigate this connection. 
In particular, a powerful technique to study the distribution of galaxies within dark matter haloes 
is the halo occupation distribution \citep[HOD, e.g.,][]{Peacock2000,Seljak2000,Scoccimarro2001,Berlind2002}.
The occupation function is defined as the average number of 
galaxies that populate a halo as a function of halo mass, $\langle {\rm N}({\rm M}_{h}) \rangle$.
In other words, the HOD is computed by the probability distribution $P(N|M_{h})$, 
which represents the probability that a halo of mass $M_{h}$ hosts $N$ galaxies with certain selected properties.
The HOD approach has shown to be useful for interpreting galaxy clustering measurements and constraining
models of galaxy formation \citep[e.g.,][]{Benson2000,Zentner2005,Zehavi2011}.
The standard HOD scheme assumes that the galaxy population depends solely on the mass of the dark matter halo,
motivated by the excursion set formalism.
However, this assumption might be inaccurate if the galaxy population within haloes depends on additional halo properties.

It is currently well established in $\Lambda$CDM simulations
that the spatial clustering of dark matter haloes depends on different halo properties
besides their mass. This dependence is commonly referred to as \textit{halo assembly bias}.
Different studies have investigated the dependence with varied halo properties such as their
formation time, accretion rate, spin, shape, velocity dispersion, concentration and anisotropy 
\citep[e.g.,][]{Sheth2004,Avila-Reese2005,Gao2005,Wechsler2006,GaoWhite2007,Faltenbacher2010,Sunayama2016,Mao2018}. 
The origin of halo assembly bias is still unclear, although different explanations have been proposed.
One possible mechanism is described by \citet{Hahn2009}, where they suggest that low mass haloes
in filaments are driven by tidal suppression of the halo growth rate in the 
vicinity of a neighbouring massive halo. Also long wavelength modes might affect the halo formation times,
and prevent the uncorrelated random walks adopted by the excursion set model \citep{Zentner2007,Dalal2008}.

If the galaxy properties closely correlate with the halo formation history, halo assembly bias might also be reflected in the 
galaxy distribution. We refer to that effect as \textit{galaxy assembly bias}
\citep[e.g.,][]{Zhu2006,Croton2007,Reed2007,Zu2008,Chaves-Montero2016,Zehavi2017}.
If this is the case, the standard HOD formalism would be incomplete and different complex
approaches should be considered \citep{Zentner2014,Hearin2016}.
However, the observational and theoretical evidence for the existence of galaxy assembly bias is still controversial and under debate.
From the observational point of view, different studies claim a detection of this
effect \citep[e.g.,][]{Yang2006,Wang2008,Wang2013,Lacerna2014,Miyatake2016,Montero2017,Tinker2017,Tojeiro2017},
while other works show small or negligible impact on galaxy properties \citep[e.g.,][]{Blanton2007,Tinker2008,Vakili2016,Dvornik2017},
or systematics in the previously claimed detections \citep[e.g.,][]{Campbell2015,Zu2017,Tinker2017B}.

A way to study the theoretical predictions for galaxy assembly bias is through the analysis of different
galaxy formation models. 
Previous reports have explored galaxy assembly bias in semi-analytic models and hydrodynamical cosmological simulations.
For instance, \citet{Croton2007} demostrate the existence of galaxy assembly bias 
in a semi-analytical model applied onto the Millennium Simulation, by comparing the two-point correlation functions from the  
model galaxies with a shuffled version, where assembly bias is erased.  
\citet{Chaves-Montero2016} do a similar analysis in the \eagle\ simulation, finding an even stronger effect.
Using galaxies in the Illustris simulation, \citet{Bray2016} detect a significant signal of galactic conformity
\citep[i.e., the correlation between star formation and colour of central galaxies and their neighbours; see][]{Kauffmann2013}, 
 a signature that has been linked to assembly bias \citep{Hearin2015,Hearin2016b}.
 
The subhalo abundance matching technique (SHAM) is another useful method to consider when discussing
galaxy assembly bias signatures. SHAM connects dark matter substructures with galaxies using a direct 
relation between a galaxy property such as its stellar mass or luminosity,
and a subhalo property like its infall subhalo mass or its maximum circular velocity 
\citep{Conroy2006,Reddick2013,Mao2015,Chaves-Montero2016,Lehmann2017,Dragomir2018}.

Signatures of galaxy assembly bias can be studied by directly exploring how the halo occupation might depend 
(or not) on different halo properties. Following \citet{Zehavi2017}, we refer
to this dependence as \textit{occupancy variation}. 
The occupancy variation was studied previously in different galaxy formation models.
\citet{Berlind2003} and \citet{Mehta2014} investigate the environmental variations of the HOD in hydrodynamical cosmological simulations, finding no significant signals of this dependence.
\citet{McEwen2018} use the age-matching catalogues 
of \citet{Hearin2013} which exhibit by construction strong galaxy assembly bias signal, to explore the dependence of HOD with environment. 
Their findings suggest this dependence exist mainly for central galaxies.
Most recently, \citet{Zehavi2017} investigate the dependence of the halo occupancy on large-scale environment and halo formation time,
by using two different semi-analytic galaxy formation models implemented in the Millennium simulation.
Their results show distinct features of occupancy variation. Central galaxies in high dense large-scale environments
are more likely to reside in lower-mass haloes. A much stronger signal is found with halo formation time, where 
early-forming low mass haloes are more likely to host a central galaxy. Furthermore, they find a reverse
trend for the satellite galaxies, with early-forming haloes containing less satellites.

In this work, we extend the investigation presented in \citet{Zehavi2017} with semi-analytical models,
to analyse the impact of the occupancy variation reflected in the
state-of-the-art hydrodynamical cosmological simulations \eagle\ and Illustris \citep{Schaye15,Vogelsberger2014}.
For this purpose, we analyse the HOD and 
its dependence on the halo formation redshift, and the large-scale environment of the haloes.
We explore the trends for three different number densities with galaxies ranked by their stellar mass,
when selecting $20\%$ of the haloes in the most and least dense environments and the $20\%$ youngest and oldest haloes.

This paper is organised as follows. In  \S~\ref{sec:sim}
we present a brief overview of the main features of the \eagle\ and Illustris simulations.
In \S~\ref{sec:definitions} we provide the definitions adopted to construct the samples
according to halo environment and formation time. Our findings are presented in \S~\ref{sec:results},
and the halo occupancy variation in the context of the stellar mass -- halo mass relation is discussed in \S~\ref{sec:SMHM}.
The main conclusions are summarised in \S~\ref{sec:conclusions}.
We discuss our large-scale environment definition in Appendix~\ref{sec:app}, and
we characterise the HOD of galaxies (stellar mass ranked) from \eagle\ and Illustris in Appendix~\ref{apend:fitsHOD}.
We analyse the occupancy variation in the context of subhaloes in Appendix~\ref{apend:NsatNsub}.

\section{The hydrodynamical simulations}\label{sec:sim}

In this section we describe the main relevant properties of the hydrodynamical cosmological simulations
\eagle\ and Illustris used in this work, and our samples selection.
We note that the cosmological parameters adopted for each simulation are different (see below).

\subsection{The \eagle\ hydrodynamical simulation}\label{sec:simEAGLE}
The \eagle\ simulation suite \citep[][]{Schaye15,Crain15} is a set of cosmological hydrodynamical simulations 
performed with a modified version of \gadget-3 code \citep[based on \gadget-2, see][]{Springel05}.
It is composed by different runs where the resolution and box sizes are varied, starting from $z=127$ up to $z\sim 0$,
and adopting the $\Lambda$CDM cosmology with parameters inferred from \cite{Planck13}
\footnote{$\Omega_{\rm m} = 0.2588$, $\Omega_\Lambda = 0.693$, $\Omega_{\rm b} = 0.0482$, $\sigma_{8} = 0.8288$, $n_{\rm s} = 0.9611$,
and $H_0 = 100\; h$ km s$^{-1}$ Mpc$^{-1}$ with $h = 0.6777$.}.
The hydrodynamical cosmological simulations uses sub-grid physics to describe the 
physical processes that are below their resolution limit. For the case of the \eagle\ suite, the sub-grid physics
is based on that developed for {\sc owls} and {\sc gimic} simulations \citep[][respectively]{Schaye2010,Crain2009}. 

Here we describe briefly the most important aspects of the sub-grid physics implemented in the \eagle\ suite.
Star formation follows the pressure-dependent Kennicutt-Schmidt relation from \citet{Schaye08}
and the metallicity-dependent density threshold of \citet{Schaye2004}.
Therefore, gas particles that fulfil the condition for star formation are converted to a
collisionless star particle stochastically with a probability which depends on the time step and the star formation rate.
Radiative cooling and heating are implemented element-by-element adopting the model described in \citet{Wiersma09a}.
UV/X-ray ionizing background is included from \citet{Haardt01} at $z=11.5$ consistent with the measurements from \citet{Planck13}.
Stellar evolution and chemical enrichment is implemented following \citet{Wiersma09b} and adopting a
\citet{Chabrier03} initial mass function in the mass range of [0.1,100]~M$_{\odot}$.
In this way, the simulation tracks the stellar mass loss and enrichment of the interstellar medium
from AGB stars, type II (core collapse) supernovae, and type Ia supernovae.
Thermal feedback from stars is implemented following the stochastic method described in \citet{DallaVecchia12}.
Black holes seeds are located in haloes more massive than 10$^{10}$~$h^{-1}$~M$_{\odot}$
and are tracked (merging and accretion) following \citet{Springel2005b,Booth09}. 
The model also includes the gas accretion onto black holes with a modified version of the Bondi-Hoyle accretion
rate described in \citet{Rosas15}. Thermal feedback from accreting black holes is also implemented stochastically \citep[see][]{Schaye15}.

The \eagle\ suite of simulations are tuned to reproduce the galaxy stellar mass function, galaxy sizes and the relation between
the stellar mass and black hole mass \citep{Shen2003,Baldry2012}. The simulation also reproduce a wide variety of observables
that are not tuned such as the specific star formation rates, the Tully-Fisher relation, the stellar luminosities
of galaxy clusters, the luminosity function and colour-magnitude diagram at $z~=~0$ \citep{Trayford2015} and 
the evolution of the galaxy stellar mass function \citep{Furlong2015}.
Furthermore, the galaxy clustering within the \eagle\ simulation was also studied in \citet{Chaves-Montero2016} and
\citet{Artale2017} finding good agreement with observations at small scales.

In this work, we use the galaxy catalogues at $z=0$ of the simulation named as L0100N1504 
from the \eagle\ suite (hereafter we refer to this run as \eagle\ \footnote{{\url http://icc.dur.ac.uk/Eagle/}, {\url http://virgo.dur.ac.uk/data.php}}),
which are available on their {\sc SQL} database \citep[see][]{McAlpine2016}.
This run consists of a periodic box of 67.77~$h^{-1}$Mpc (100~Mpc) side,
which initially contain 1504$^3$ of gas and dark matter particles, with masses of 
$m_{\rm gas} = 1.23\times10^6 h^{-1} {\rm M}_{\odot}$ and $m_{\rm DM} = 6.57\times10^6 h^{-1} {\rm M}_{\odot}$.

\subsection{The Illustris hydrodynamical simulation}\label{sec:simIllustris}
The Illustris Project \citep{Vogelsberger2014,Genel2014} is a set of cosmological hydrodynamical simulations 
of periodic ($75h^{-1}$Mpc)$^3$ volume that was performed with the {\sc arepo} moving-mesh code of 
\citet{Springel2010}.
The simulations were run following the evolution of dark matter, gas and star particles
with three different resolution levels from $z=127$ to $z=0$ within a $\Lambda$CDM cosmological scenario\footnote{$\Omega_{\rm m} = 0.2726$, $\Omega_{\Lambda} = 0.7274$, $\Omega_{\rm b} = 0.0456$, $\sigma_{8} = 0.809$, $n_{\rm s} = 0.963$, 
and $H_{0} = 100~h~{\rm km s^{-1} Mpc^{-1}}$ with $h = 0.704$.}. 

The Illustris simulations also include sub-grid models for those physical processes out of the resolution limits of the simulation
\citep[for further details see,][]{Vogelsberger2013}.
Star formation is implemented following \citet{Springel2003}, where gas particles are promoted to star 
stochastically when their density is above a certain limit within a star formation time-scale.
Stellar evolution and chemical enrichment are implemented adopting the \citet{Chabrier03} initial mass function in the 
same mass range than in \eagle. Therefore, the model follows the mass loss and chemical contribution into the interstellar medium from AGB stars,
core collapse supernovae and Type Ia supernovae.
Radiative cooling is implemented through spatially uniform time-dependent UVB and metal-line cooling
based on {\sc cloudy} \citep[see][]{Vogelsberger2013}.
Stellar winds and supernovae feedback are implemented following an adapted model from \citet{Springel2003}.
The seed of black holes are located in dark matter haloes with masses above $5\times 10^{10} h^{-1}$~M$_{\odot}$, 
and active galactic nuclei feedback implementation follow previous studies \citep{DiMatteo2005,Springel2005b}.

In Illustris the free parameters from the sub-grid models were tuned to reproduce the cosmic star-formation
rate density and the stellar mass function at $z=0$ in a smaller volume of $\sim 25 h^{-1}$Mpc (35.5~Mpc) of a side. 
This simulation has also shown to reproduce different observed properties that are not tuned such as the distribution of 
the galaxy morphologies \citep{Snyder2015}, 
the build-up of galactic mass, the evolution of galaxy specific star formation rates up to $z = 8$ \citep{Genel2014}
and the colours of satellites \citep{Sales2015}.
However, some observational properties are not well reproduced such as 
the galaxy stellar mass function for the 75~$h^{-1}$Mpc side box,
finding at $z \sim 0$ an excess of galaxies at high 
and low stellar masses \citep[see][]{Vogelsberger2014}.

The Illustris project provides open access to their database\footnote{http://www.illustris-project.org},
where it is possible to obtain information about the particle and galaxy catalogues at different redshifts \citep[see,][]{Nelson2015}.
In this work we use Illustris-1, the highest resolution simulation run which initially contains 1820$^3$ gas and dark matter particles
 with masses of $m_{\rm gas} = 1.13 \times 10^{6}$~$h^{-1}$~M$_{\odot}$ and $m_{\rm DM} = 4.44 \times 10^{6}$~$h^{-1}$~M$_{\odot}$.
An updated version of the physical model implemented was recently published improving some of the issues mentioned above.
This new version, known as IllustrisTNG \citep[][]{Springel2017,Pillepich2018}, includes the effect of seed magnetic fields,
an updated model for galactic winds and a new implementation of supermassive black holes kinetic feedback. 
Once publicly available, it would be interesting to revisit our analysis using the improved
IllustrisTNG simulation.

 \begin{figure}
 \centering
 \includegraphics[width=0.45\textwidth]{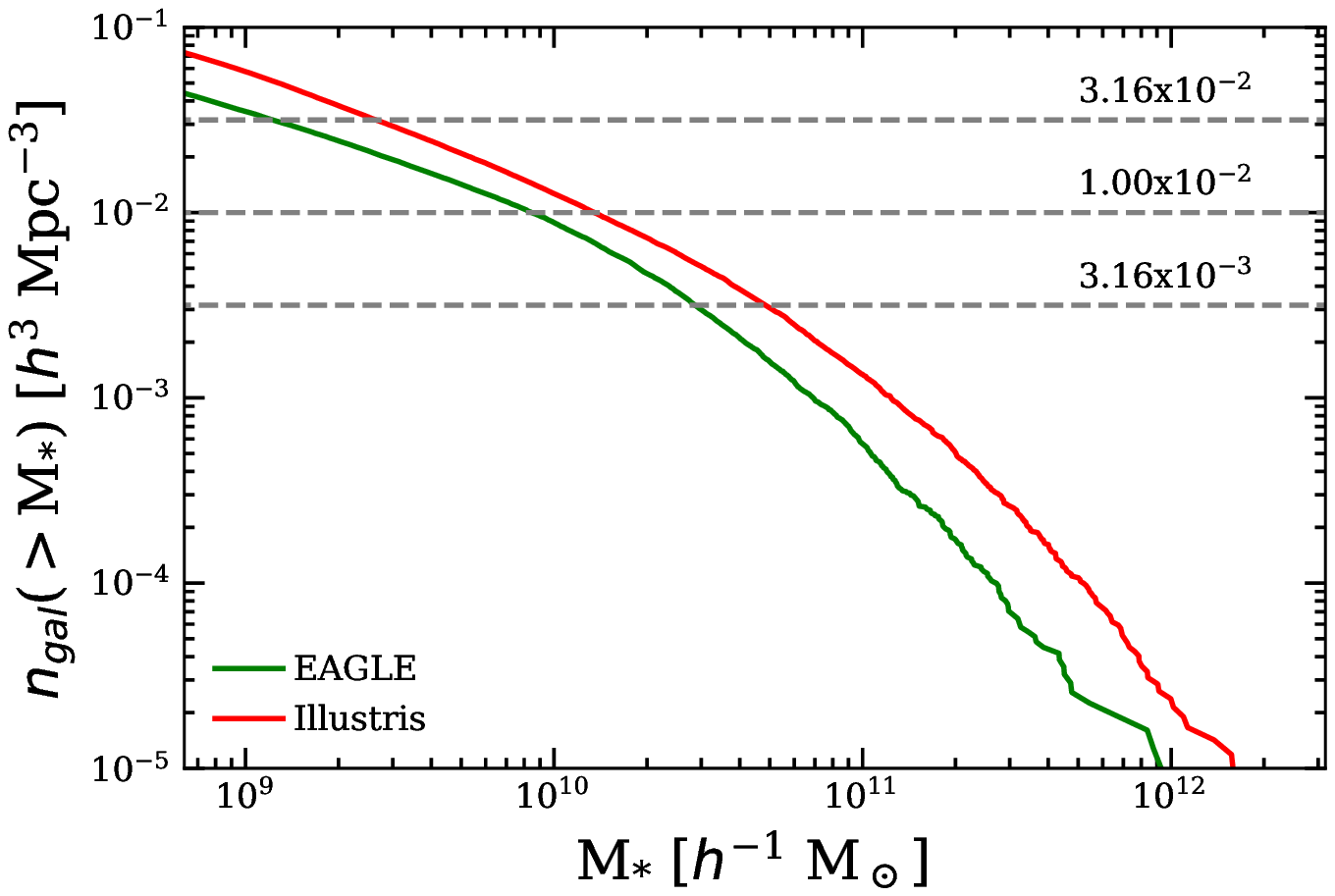}
 \includegraphics[width=0.45\textwidth]{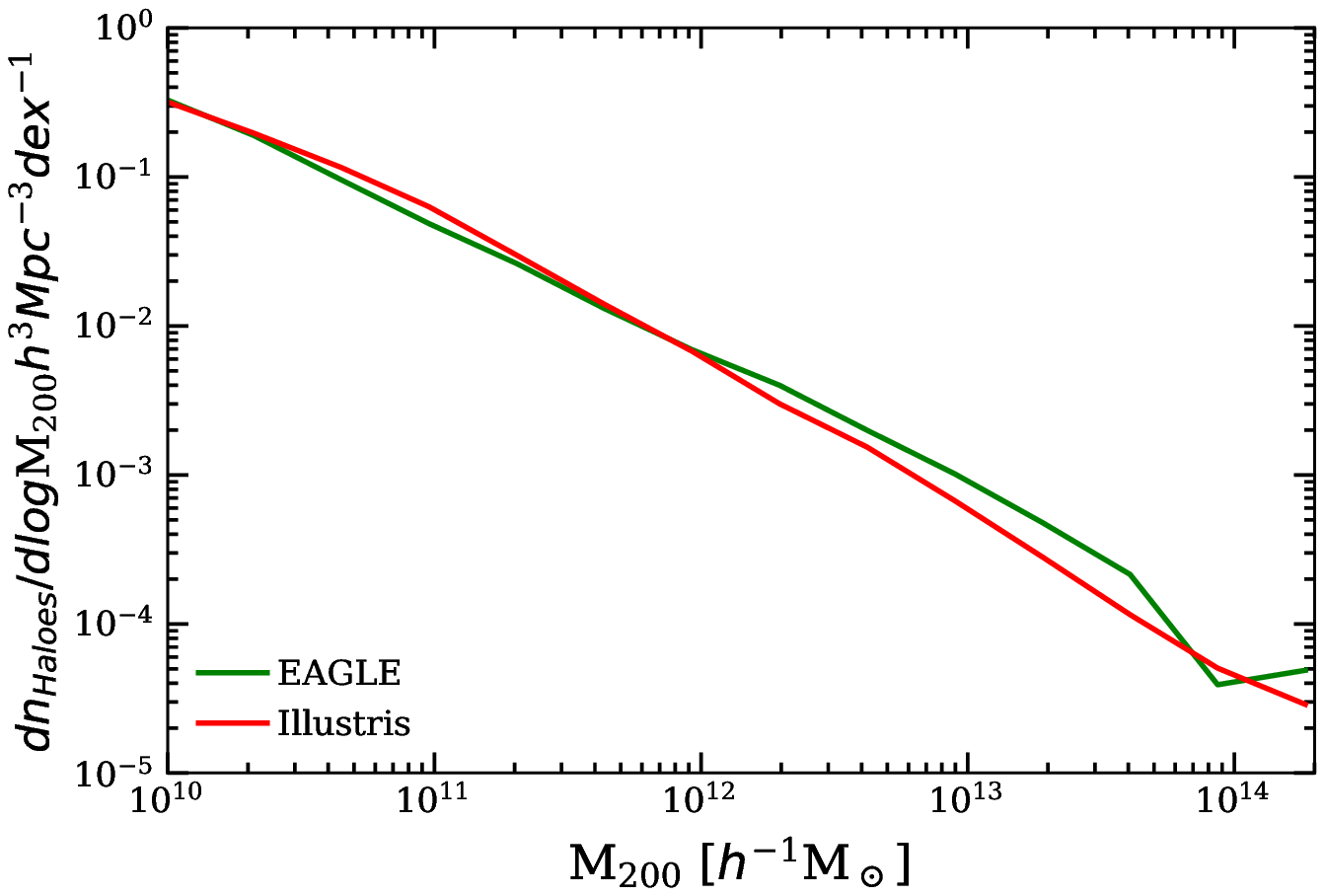}
 \caption{\textit{Top panel:} The cumulative stellar mass function of \eagle\ (green line) and Illustris (red line).
 The dashed lines represent the galaxy number densities selected to estimate the HOD corresponding to $n=(0.0316,0.01,0.00316)$~$h^3$~Mpc$^{-3}$.
 \textit{Bottom panel:} Halo mass functions of \eagle\ (green line) and Illustris (red line). We use M$_{200}$ as the halo mass.}
 \label{fig:CSMF}
 \end{figure}
 
\subsection{Selection of galaxies and dark matter haloes}\label{sec:sim-selection}

In both simulations the dark matter haloes are identified using the \textit{Friends-of-Friends} (FoF) method on the dark matter only
simulation with a linking length of 0.2 times the mean interparticle separation \citep{Davis1985}.
Gravitationally bound substructures (or subhaloes) within haloes are identified using the 
SUBFIND algorithm \citep{Springel2001} in both simulations. 
The SUBFIND algorithm uses the gas, stellar and dark matter particles assigned to each FoF halo \citep{Schaye15}.
Galaxies are associated with the baryonic component within each subhalo.
Central and satellite galaxies are identified and provided by the simulation teams.
In \eagle\/, the central galaxy resides in the subhalo that contains the most bound dark matter particle
in the halo, while the remaining subhaloes host the satellite galaxies \citep[see][]{McAlpine2016}.
On the other hand, in the Illustris simulation, central and satellite galaxies are labelled by the mass of the subhaloes, where
the most massive one hosts the central galaxy.

We note that small subhaloes may not contain stellar and/or gas particles even at $z\sim 0$.
We use the full sample of the subhaloes to define the environment of the haloes,
irrespective of whether they contain a galaxy or not (see \S~\ref{sec:defEnv} for further details).
In this work we use the following information provided by the database of each simulation: the stellar mass of the galaxies (M$_{*}$),
the mass of the halo (M$_{\rm 200}$) defined as the total mass (i.e., the sum over dark matter and baryonic mass)
enclosed within a sphere with a density equal to 200 times the critical density,
and the co-moving positions of the galaxies\footnote{The co-moving position of a galaxy is defined as
the minimum of the gravitational potential, as defined by the most bound particle.}.

We also use the information supplied by the subhalo merger trees to define the formation time of the dark matter
haloes (see \S~\ref{sec:defFormTime}).
In the case of \eagle\, the descendant subhaloes are identified using the {\sc D-Trees} algorithm \citep[][]{Jiang2014,Qu2017}, which traces the subhaloes
using the most bound particles of any species.
The main progenitor branch is defined as the progenitor with the largest mass in each redshift.
For Illustris, the merger trees are computed using three different algorithms: {\sc SubLink} \citep{RodriguezGomez2015},
{\sc LHaloTree} \citep{Springel2005b} and {\sc Consistent-Trees} using {\sc Rockstar} \citep{Behroozi2013}.
Here we use the subhalo merger tree computed with {\sc SubLink}, available in the database.
We also use the main branch defined as the one with the most massive history.

We present the cumulative comoving number density of galaxies for \eagle\ and Illustris in the
top panel of Fig.~\ref{fig:CSMF} (green and red lines, respectively).
The Illustris simulation contains
more massive galaxies than the \eagle\ simulation, which is mainly explained by the 
differences in the sub-grid model for feedback from star-formation \citep[see][]{Schaye15}. 
In the bottom panel of Fig.~\ref{fig:CSMF} we present the halo mass functions for both simulations. We generally 
find a good agreement,
while at high halo masses, haloes in \eagle\ are more numerous than in Illustris. This is consistent with sample variance
expected from the size of the boxes used \citep{Artale2017}.

To analyse the occupancy variation, we adopt criteria to select dark matter haloes according to their
formation time or environment. We use three different number density samples,
ranking galaxies by stellar mass, corresponding to n $ = (0.0316,0.01,0.00316)$~$h^{3}$~Mpc$^{-3}$.
These are denoted in the top panel of Fig.~\ref{fig:CSMF} by the dashed lines
(see Sec.~\ref{subsec:HOD}, for further discussion regarding the selected samples).


\section{Halo samples}\label{sec:definitions}

In this work we investigate the occupancy variation of the HOD with halo formation time and large-scale environment.
We first select the dark matter haloes with  M$_{200}>10^{10}~h^{-1} {\rm M}_{\odot}$ and their respective subhaloes at $z=0$.
We note that the halo mass functions are slightly different between \eagle\ and Illustris (see \S~\ref{subsec:HOD} and Fig.~\ref{fig:CSMF}).
Those differences need to be accounted for when making a direct comparison of the HOD findings from each simulation.

In the case of \eagle\ simulation, the mean number density of selected haloes that fulfill the halo mass condition is $\sim 0.16~h^{3} {\rm Mpc}^{-3}$, while for Illustris it is $\sim 0.2~h^{3} {\rm Mpc}^{-3}$.
It is important to consider the cosmic variance among the two simulations.
\citet{Artale2017} studied in detail the impact of cosmic variance on the two-point galaxy correlation 
function from the two simulations.
The halo mass cut is adopted conforming to the resolution limits of the simulations, in order to have more than $\sim 1500$ dark matter
particles within each halo. This is valid for both simulations.
In this section, we present the definitions implemented for the halo formation time and the large-scale environment and how we split the halo populations by these properties in order to study the occupancy variation.

\begin{figure*}
  \centering
  \includegraphics[width=0.45\textwidth]{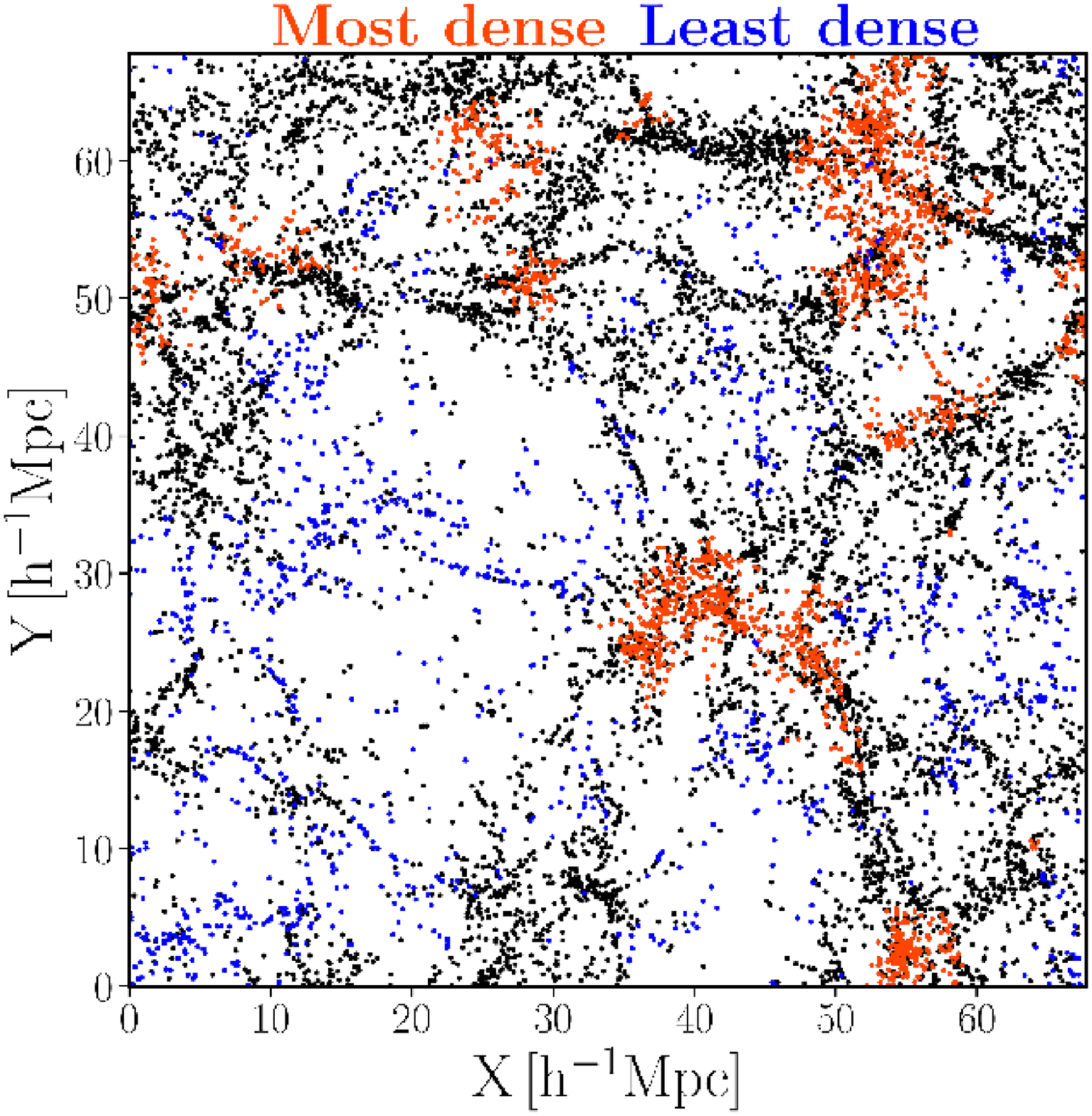}
  \includegraphics[width=0.468\textwidth]{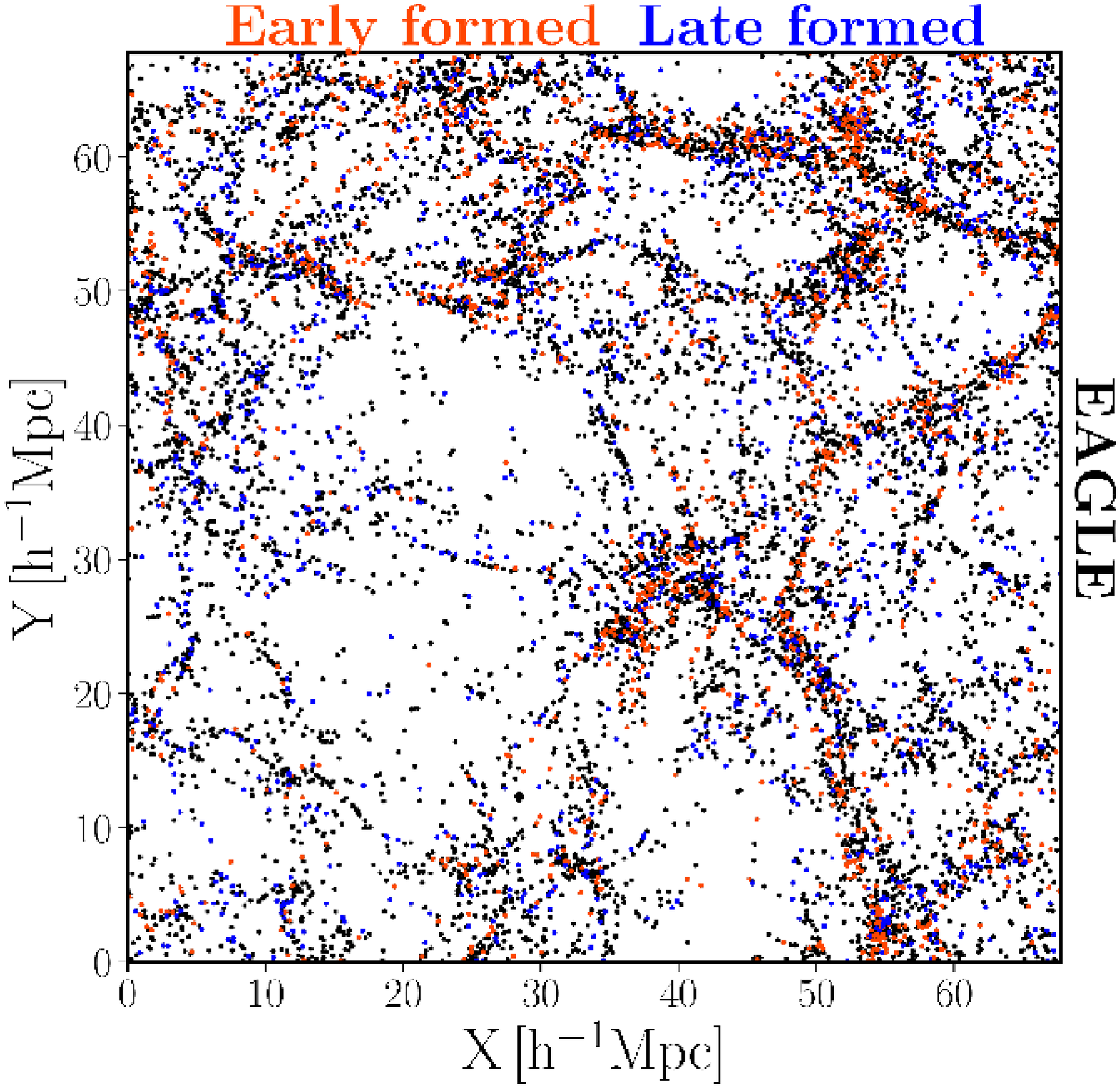}
  \includegraphics[width=0.45\textwidth]{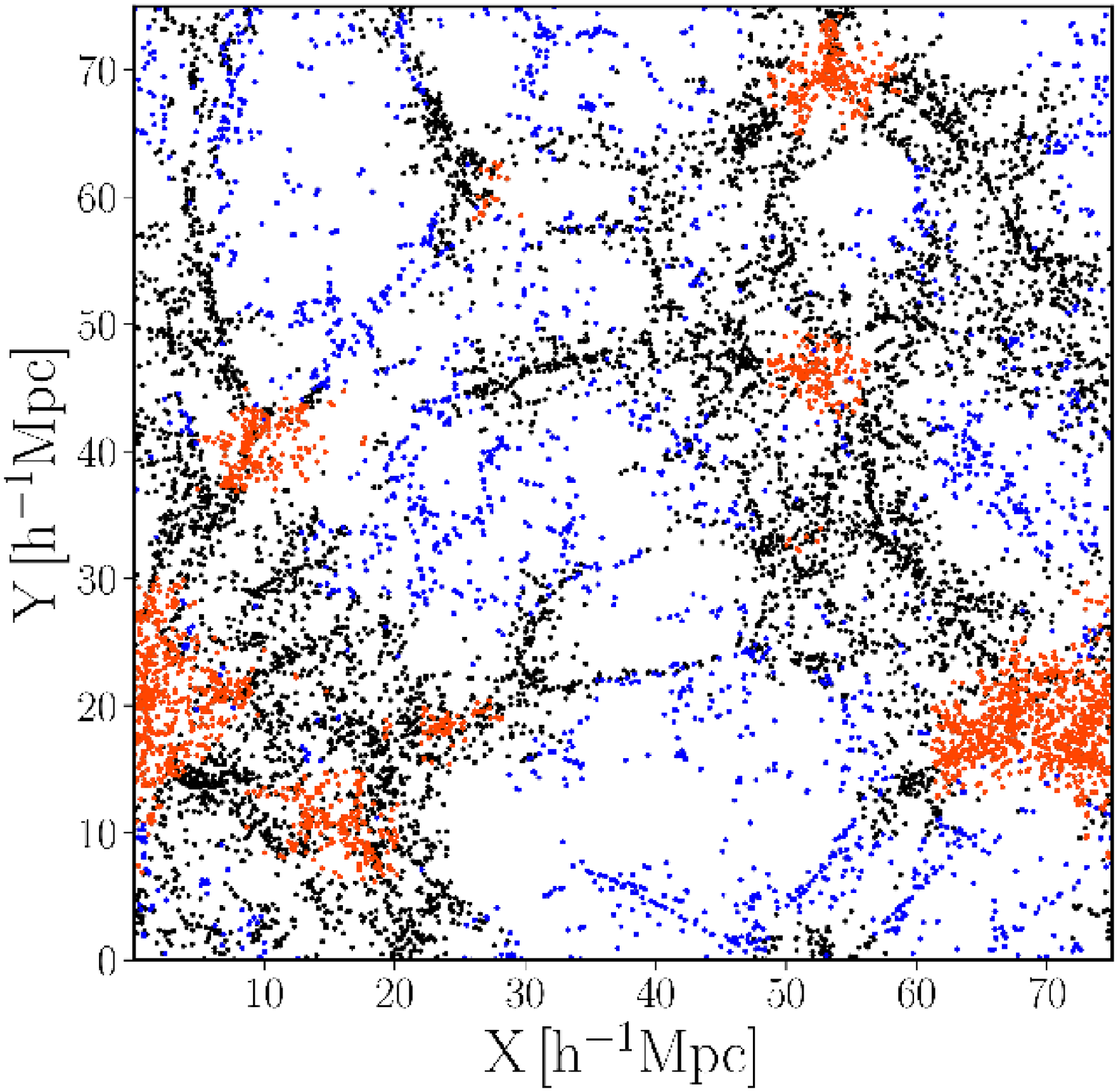}
  \includegraphics[width=0.468\textwidth]{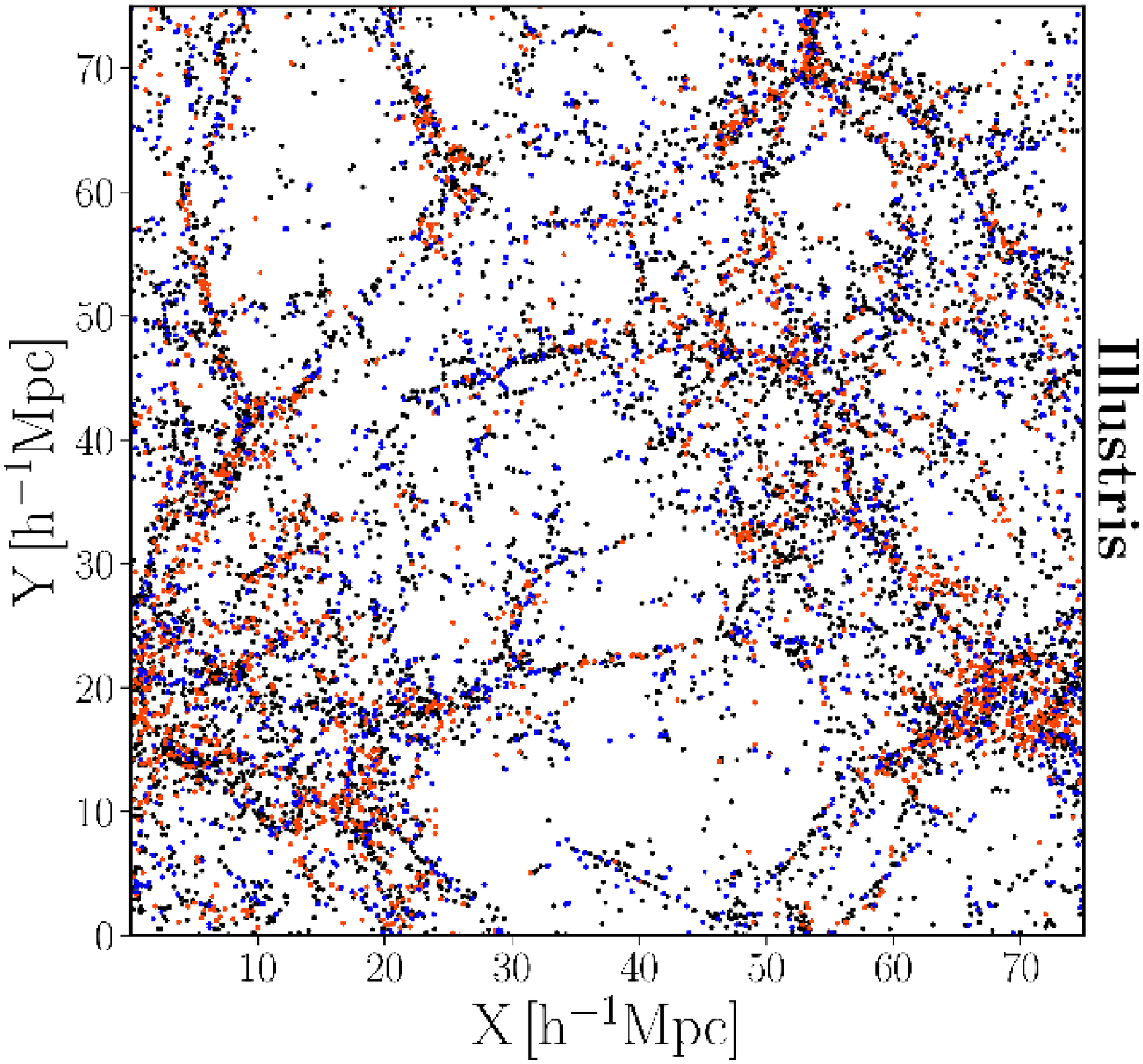} 
  \caption{
  Spatial distribution of the dark matter haloes split by environment (\textit{left panel})
  and formation time (\textit{right panel}) for haloes with masses $M_{200} > 10^{10}  h^{-1} {\rm M}_{\odot}$.
  We present a slice of 67.7x67.7x10~$h^{-1}$~Mpc of the \eagle\ simulation
  (\textit{top panels}), and a slice of 74.9x74.9x10 $h^{-1}$~Mpc for Illustris simulation (\textit{bottom panels}).
  In each case, the x and y ranges shown correspond to the maximum comoving length of each simulation.
  In the left panels, red and blue dots represent 20\% of the haloes in the most and least dense environments, respectively.  
  In the right panels, red dots represent the haloes that belong to the 20\% earliest-formed (oldest) haloes, while
  the blue dots are the 20\% latest-formed (youngest) haloes.
  Black points show haloes not belonging to any of the two selected populations. See \S~\ref{sec:definitions} for more details.
  }
  \label{fig:SpaceDistrib}
\end{figure*}

\subsection{Selection by large-scale environment}\label{sec:defEnv}

We investigate the occupancy variation due to the large-scale environment.  
We define the large-scale environment of each dark matter halo
by counting all the subhaloes within a sphere of 5~$h^{-1}$Mpc radius
(excluding those that belong to the same halo), divided by the volume of the sphere, 
and adopting periodic boundary conditions (we denote this environment measure as $n_{5{\rm Mpc}/h}$). 

We note that we are counting subhaloes and not galaxies, since some subhaloes
may not contain gas and/or stellar particles (see \S~\ref{sec:sim-selection}). 
For both simulations we use all the subhaloes provided by each database that belong to the selected haloes \citep[see][for further details]{McAlpine2016,Nelson2015}.  
Hence, the environment definition adopted depends on the resolution limit of each simulation. 
While this should be taken into account in future comparisons, we do not expect this to significantly impact our main results.

We normalize this quantity by the number density of subhaloes within haloes of M$_{200}>10^{10}~h^{-1} {\rm M}_{\odot}$ (referred to $n_{avg}$).
We tested different radii for the sphere (see~Appendix~\ref{sec:app} for more details).
We select a radius of 5~$h^{-1}$Mpc to quantify the environment as a compromise between covering volumes larger than those
of the biggest haloes, but also taking into account the size of the simulated box.
The test with other radii gives us similar results, making our findings robust.

Similar to the method implemented in \citet{Zehavi2017}, to classify the dark matter haloes according
to their environment, we display the halo masses in bins of $\sim 0.24$~dex
and select those with the 20\% most and least dense environment in each halo mass bin. 
We define environment cuts separately in bins of halo mass to factor out the halo mass function dependence on environment.
Therefore, with this approach it is possible to compare the differences in the HOD at fixed halo mass and different environments.  
Additionally, all the samples will have, by definition, the same halo mass function.

We show the spatial distribution of the dark matter haloes for a slice of the \eagle\ (top) and Illustris (bottom) simulations 
in the left hand of Fig.~\ref{fig:SpaceDistrib}, 
split by those in the most and least dense environments marked by red and blue points, respectively.
Black points show those haloes in the slice not included in these selections. 
As expected, we find that the haloes from each population map different regions of the large-scale structure, where
the haloes from the high density population are distributed in more compact and clustered regions than the haloes from the low dense population.

The left panel of Fig.~\ref{fig:DefEnv} show the selection criteria of the \eagle\ haloes for the most and least dense environment (red and blue points).
Black points correspond to those haloes that do not belong to the selected populations.
We find that the density cut selected to define the haloes in the most and least dense environments gradually increases
with halo mass, reflecting the well-known dependence of the halo mass function on environment.
The median value also shows an increase with halo mass as expected.

\subsection{Selection by halo formation time}\label{sec:defFormTime}

To classify the haloes according to their assembly history, we compute the formation time of the haloes ($z_{\rm form}$), 
as the redshift at which for the first time half of present-day halo mass has been accreted into a single subhalo. For this purpose, we 
use the subhalo merger trees of each simulation to track the mass since $z=6$.
For all the haloes that at $z=0$ have a mass M$_{200}>10^{10}~h^{-1} {\rm M}_{\odot}$, we follow 
the progenitors of the main branch.
We compute the formation redshift using a spline interpolation of the masses.

To classify the haloes, we identify the 20\% early-formed (old haloes) and the 20\% late-formed (young haloes),
in fixed bins of 0.24~dex of halo mass.
The spatial distribution of the dark matter haloes for a slice of each simulation
is shown in the right side of Fig.~\ref{fig:SpaceDistrib} (\eagle\ on top, and Illustris on bottom).
Red points correspond to the dark matter haloes that belong to the sample of 20\% early-formed, while 
blue points correspond to the 20\% late-formed.
We find that the population of early and late formed haloes map different regions of the cosmic web, and
the spatial distribution is different to the one split by environment.
This suggests that the two methods to classify halo populations are not strongly correlated (see more below in this section).

\begin{figure*}
 \centering
 \includegraphics[width=0.475\textwidth]{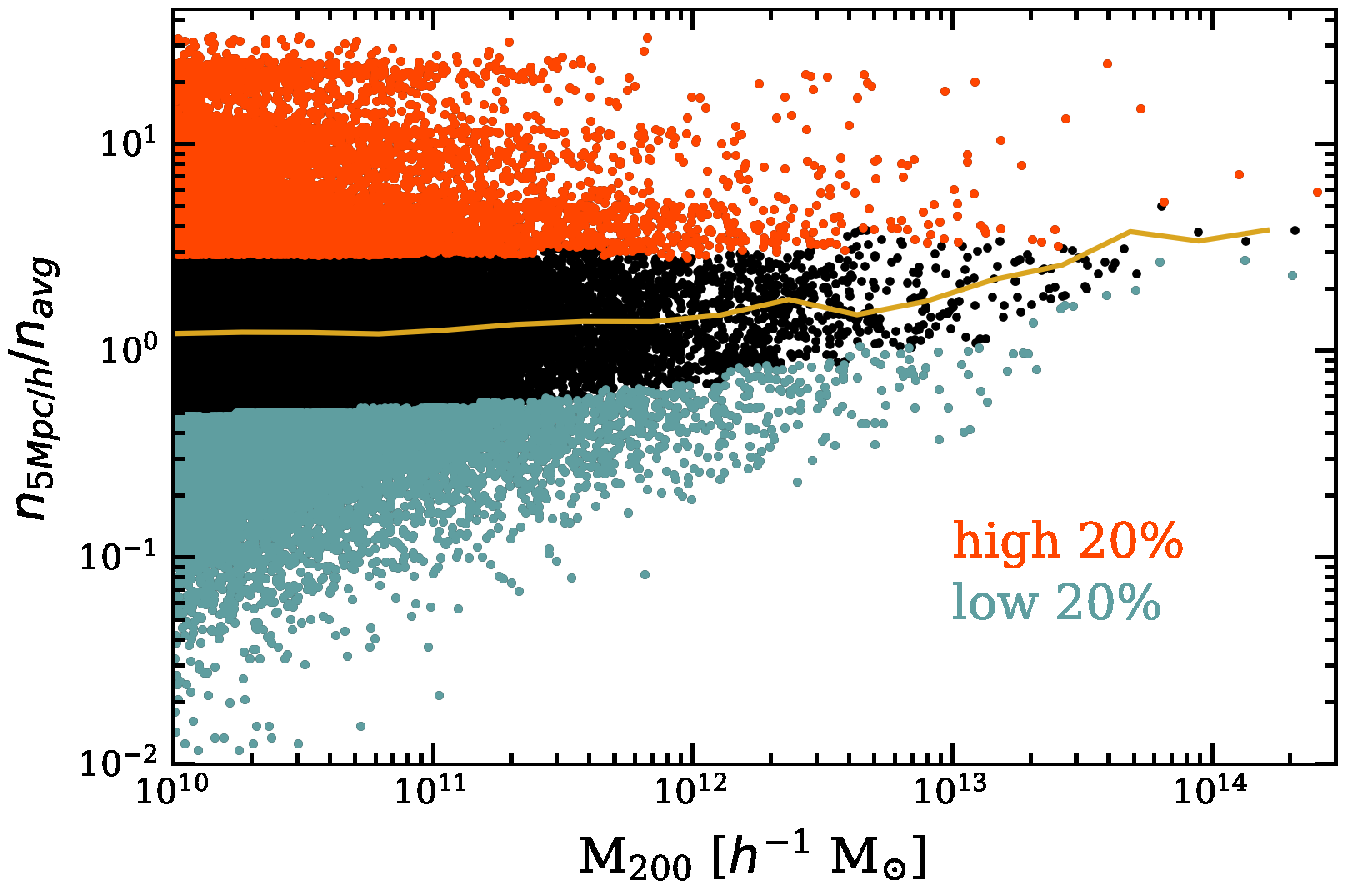}
 \includegraphics[width=0.45\textwidth]{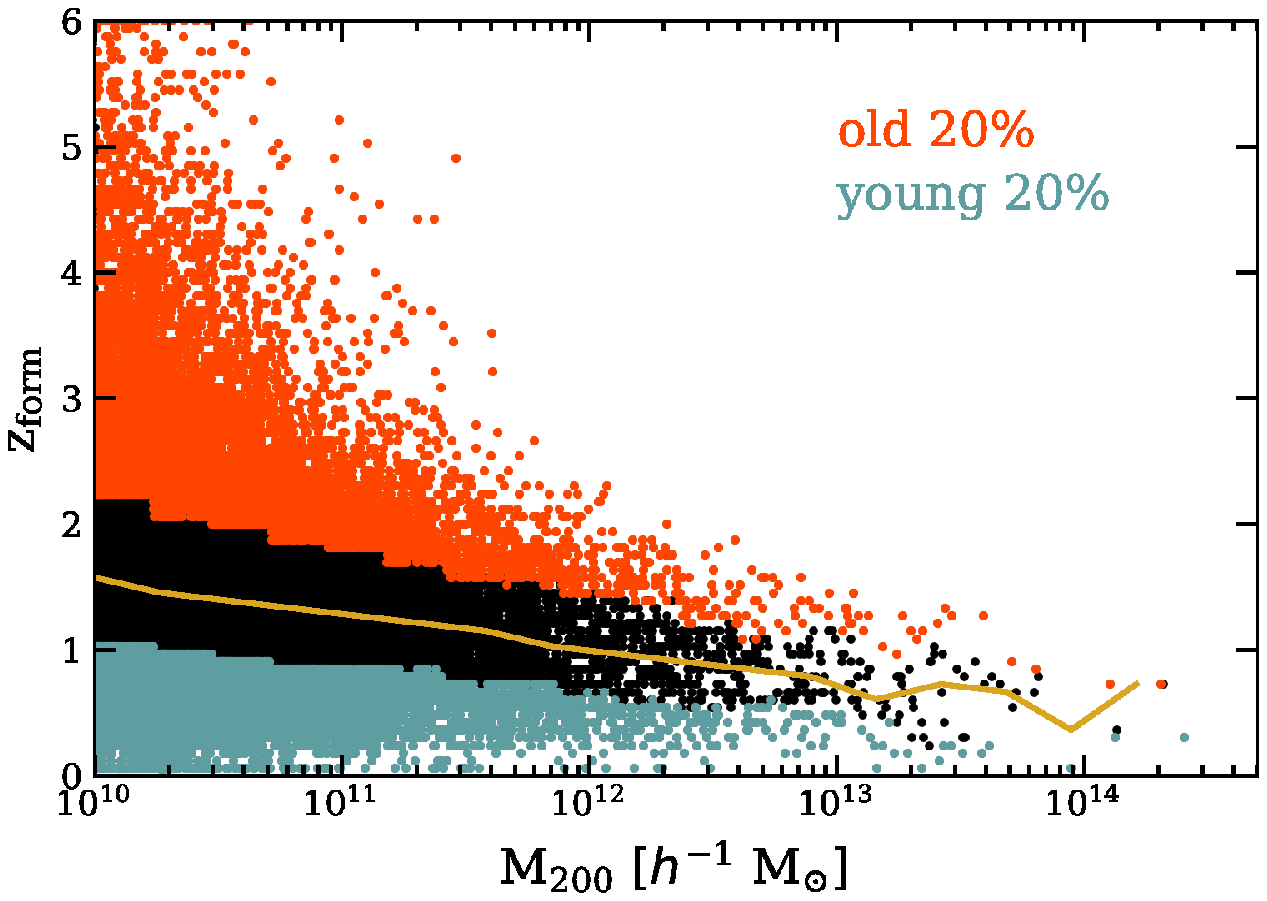}
\caption{
 \textit{Left panel:} The distribution of large-scale environment with halo mass,
 showing the different environment regions used in this work.
 We show the dark matter haloes with $M_{200} > 10^{10} h^{-1} {\rm M}_{\odot}$ from the \eagle\ simulation. 
 We split the population into 20\% highest (red points) and lowest (blue points) density environments for each
 halo mass bin. Black points are the haloes that do not belong to any of the selected samples.
 The yellow line represents the median large-scale environment for each halo mass bin. 
\textit{Right panel:} The same as the left panel, but for the formation time of the haloes ($z_{\rm form}$), 
as a function of halo mass for \eagle\, split by the 20\% latest-formed 
  (blue) and earliest-formed (red) haloes for each halo mass bin.
  }
 \label{fig:DefEnv}
\end{figure*}

The right panel of Fig.~\ref{fig:DefEnv} presents the corresponding selection criteria for the \eagle\ haloes for
the 20\% early-formed and late-formed haloes (red and blue points). Black points
are those haloes that do not belong to any of these populations, while the 
yellow line is the median of the formation time for each halo mass bin.
The population of early-formed haloes show a wide range of formation redshifts between $z \sim 1-6$ 
as function of halo mass, while late-formed haloes are generally below $z \sim 1$ for all halo masses.
The median formation time decreases with increasing halo mass, reflecting the late formation time of massive haloes.

As already mentioned, the visual comparison of the haloes' spatial distribution in Fig.~\ref{fig:SpaceDistrib} 
shows that each criterion roughly selects a distinct distribution,
where haloes selected by age and environment are not strongly correlated.
To quantify how independent these subsamples are, we compute the fraction of haloes that are in common in
young/low-density and old/high-density selections.
We find that aproximately 25\% -- 30\% of the young (old) subsample of haloes belong also to low-density (high-density) haloes
for \eagle\ and Illustris (this corresponds to $\sim~5\%$ of the 
total dark matter haloes with $M_{200}>10^{10}~h^{-1}~{\rm M}_{\odot}$).

\begin{figure}
   \centering
   \includegraphics[width=0.45\textwidth]{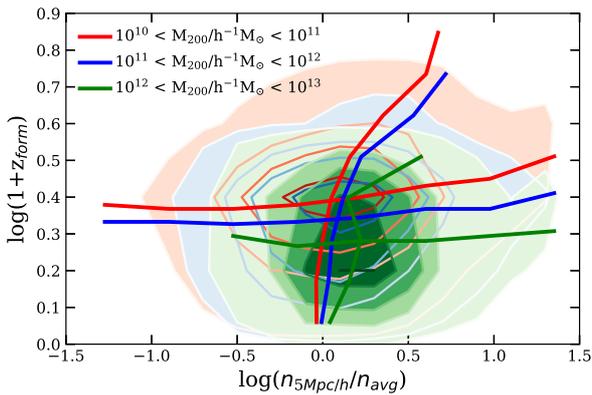}
   \caption{Joint distribution of the large-scale environment ($n_{5{\rm Mpc}/h} / n_{avg}$) and formation time ($1+z_{\rm form}$)
   for dark matter haloes of the \eagle\ sample.
   The contours correspond to enclosing 20, 50, 70, 95 per cent of the sample for three different
   halo mass ranges: $10^{10}-10^{11} h^{-1} {\rm M}_{\odot}$ (red), 
   $10^{11}-10^{12} h^{-1} {\rm M}_{\odot}$ (blue),
   $10^{12}-10^{13} h^{-1} {\rm M}_{\odot}$ (green). The roughly vertical lines represent the median values of the environment at 
   each formation redshift and halo mass bin, while
   horizontal lines correspond to the median values of the formation redshift as a function of the large-scale environment.}
\label{fig:corr-Dens-FormTime}
 \end{figure}

We also analyse the differences in halo selection for the full halo sample.
Fig.~\ref{fig:corr-Dens-FormTime} shows the correlation of 
formation time with the large-scale environment in \eagle\ for three different mass bins 
\citep[similar to the analysis presented in Fig.~2 of][]{Zehavi2017}.
For each halo mass bin, we show the median values of the environment as a function of formation redshift (vertical lines)
and the median values of the formation time as a function of large-scale environment (horizontal lines). 
We find a slight trend of early-formed haloes residing in denser environments, as indicated by the slanted
lines.
However, overall we find a broad range of halo formation times and densities, suggesting  
that there is little correlation between these quantities.
It is thus interesting to investigate the occupancy variation of these two largely independent quantities.

\section{The occupancy variation}
\label{sec:results}

In this section we present our results for the occupancy variation with large-scale environment and halo age.
We first compare the HOD of \eagle\ and Illustris for three samples with different galaxy number densities
and analyse the differences in the central and satellite galaxy populations.
We then study how the HOD varies with the large-scale environment and halo formation time. 
We clarify that while each selected subset of haloes represent 20\% of the haloes at $z=0$ with
M$_{200} > 10^{10} h^{-1} {\rm M}_{\odot}$,
the associated galaxies do not necessarily make up the same fraction.

\subsection{The halo occupation distribution}\label{subsec:HOD}

We first compare the HOD of \eagle\ and Illustris for fixed number density samples ranked by galaxy stellar mass.
This approach has shown to be a preferred alternative to 
fixed stellar mass samples since it is not affected by systematic shifts in the stellar mass estimates
and model calibrations \citep[e.g.,][]{Padilla2011,Contreras2013,Leja2013,Mitchell2013,Mundy2015,Torrey2015}.

The cumulative comoving number density of galaxies ranked as a function of stellar mass with the 
three cuts adopted was shown already in Fig.~\ref{fig:CSMF}.
The number densities selected are n$ = (0.0316,0.01,0.00316)$~$h^{3}$~Mpc$^{-3}$ for both simulations.
From Fig.~\ref{fig:CSMF} it is evident that each number density chosen corresponds  to a different 
stellar mass cut for each simulation, reflecting the variations between the galaxy formation models
and the model tuning adopted.

\begin{figure*}
 \centering
 \includegraphics[width=0.45\textwidth]{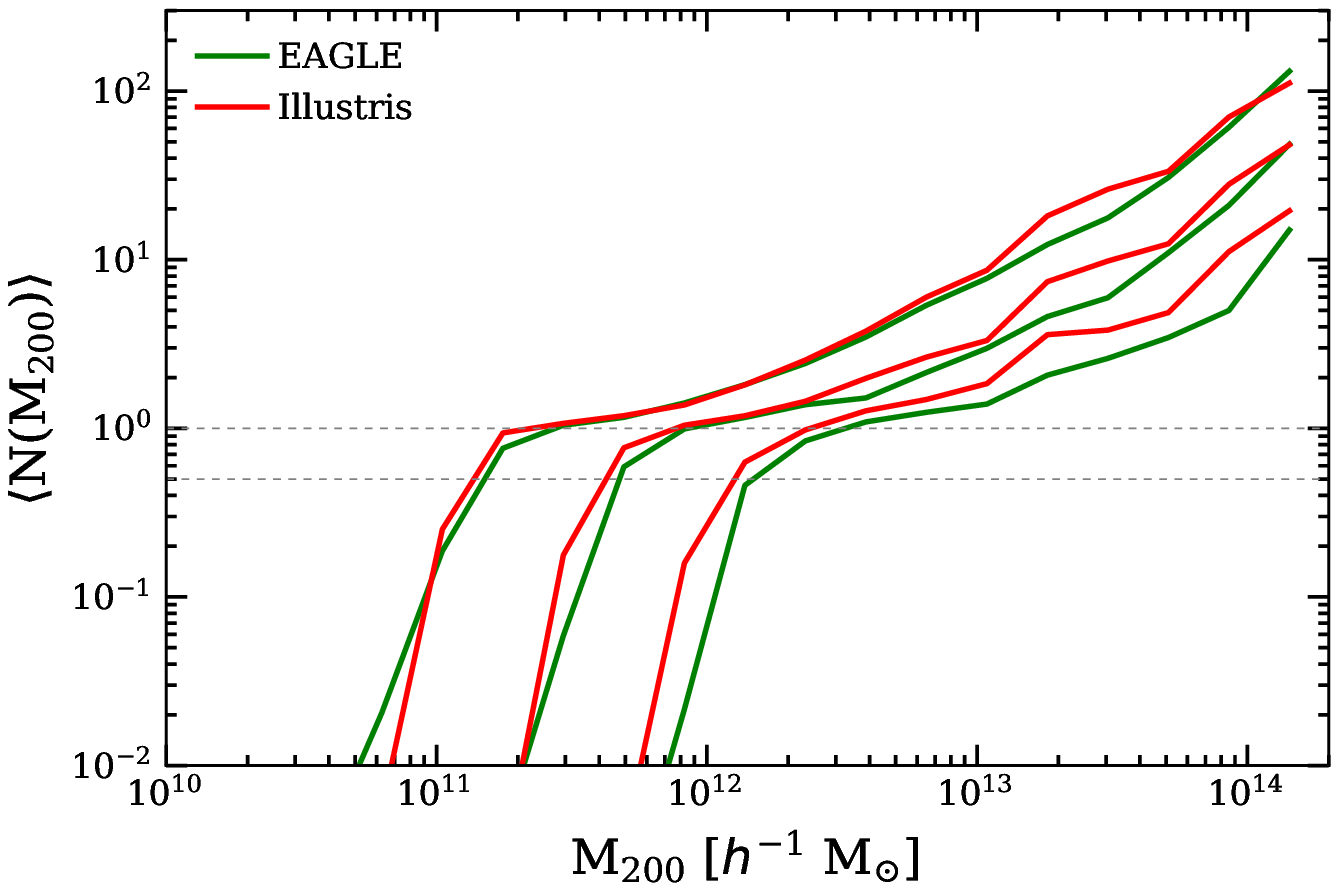}
 \includegraphics[width=0.45\textwidth]{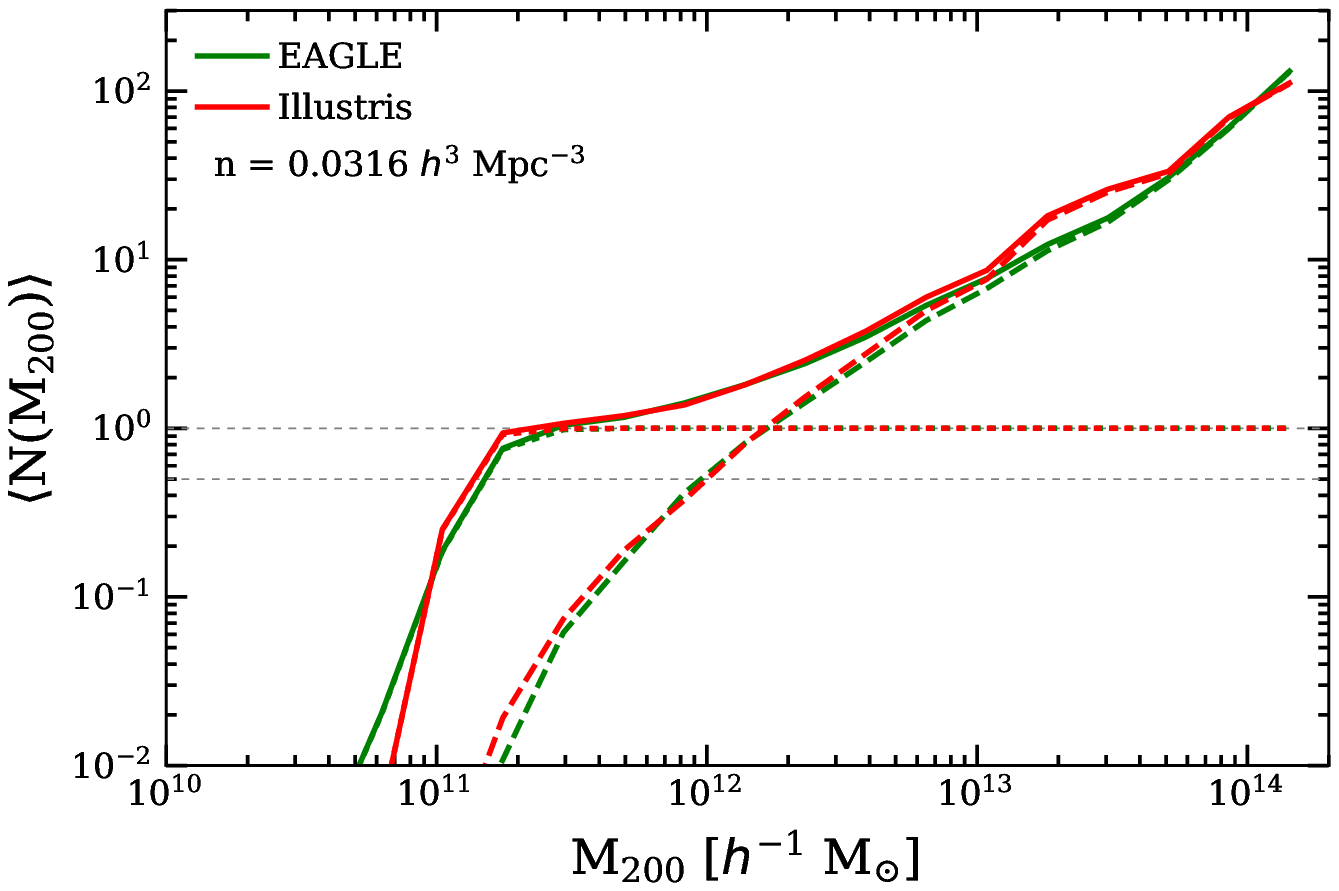}
 \caption{\textit{Left panel:} The halo occupation functions of \eagle\ galaxies (green lines) and Illustris (red lines) for the three number density cuts of $n =$ 0.0316, 0.01, 0.00316~$h^3$Mpc$^{-3}$, going from left to right.
 \textit{Right panel:} The halo occupation functions of \eagle\ and Illustris galaxies for the
 $n =$ 0.0316~$h^3$Mpc$^{-3}$ sample, split by central (dotted lines) and satellite galaxies (dashed lines).
 The horizontal grey dotted lines can be used as reference to visually estimate $M_{1}$ and $M_{\rm min}$ (see Appendix~\ref{apend:fitsHOD}).}
 \label{fig:HOD-ND}
 \end{figure*}

The left panel of Fig.~\ref{fig:HOD-ND} presents the HODs obtained from \eagle\ (green lines) and Illustris (red lines) 
for the three number densities.
We find that the mean occupation of haloes in Illustris is shifted towards lower halo masses relative to \eagle. 
This can be explained by the differences in halo mass functions seen in Fig.~\ref{fig:CSMF} 
(\textit{bottom panel}). Since \eagle\ contains a larger number of massive dark matter
haloes relative to Illustris, this translates to a shift in the occupation of the dark matter haloes.
We also show the HOD split by central and satellite galaxies in the right panel of Fig.~\ref{fig:HOD-ND}, for the galaxy samples
of \eagle\ and Illustris at n$=0.0316~$~$h^3$~Mpc$^{-3}$ finding similar results for both simulations. 
We provide parametrized fits to the HOD from \eagle\ and Illustris in Appendix~\ref{apend:fitsHOD}.
This may be useful for contrasting with different galaxy formation
models or observational results in the future.

As stated earlier, we investigate the HOD in \eagle\ and Illustris using the 
masses of dark matter haloes from the hydrodynamical simulations.
However, we note that HOD analyses are typically based on dark matter only simulations (DMO).
It is well known that baryons modify the properties of their dark matter hosts \citep[e.g.,][]{Pedrosa2010,Tissera2010,DiCintio2014,Schaller2015,Dutton2016}. Hence,
the HOD might present differences when using the halo masses from the hydrodynamical and/or DMO counterpart.
\citet{Chaves-Montero2016} find only negligible impact in the two-point correlation function of central galaxies 
when the sample is selected using the maximum circular velocity from the full \eagle\ simulation or their DMO counterpart.
Therefore, we expect this to only have a minor impact on our results.

In the following sections, we compare the occupancy variations
of Illustris and \eagle\ galaxies for the three number densities presented, with
environment and halo formation time.

\subsection{The HOD dependence on environment}\label{sec:HODenv}

\begin{figure*}
 \centering
 \includegraphics[width=0.45\textwidth]{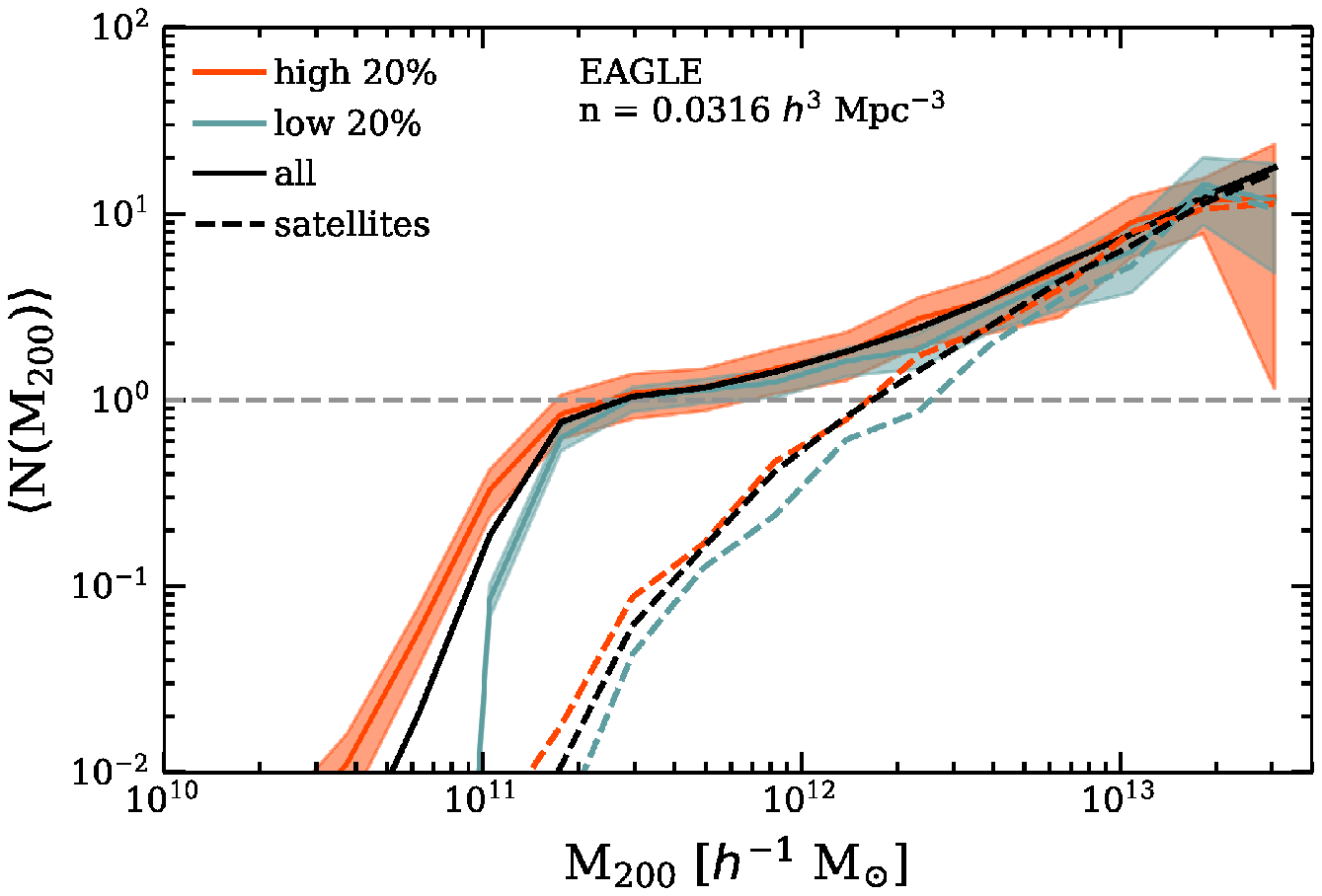}
 \includegraphics[width=0.45\textwidth]{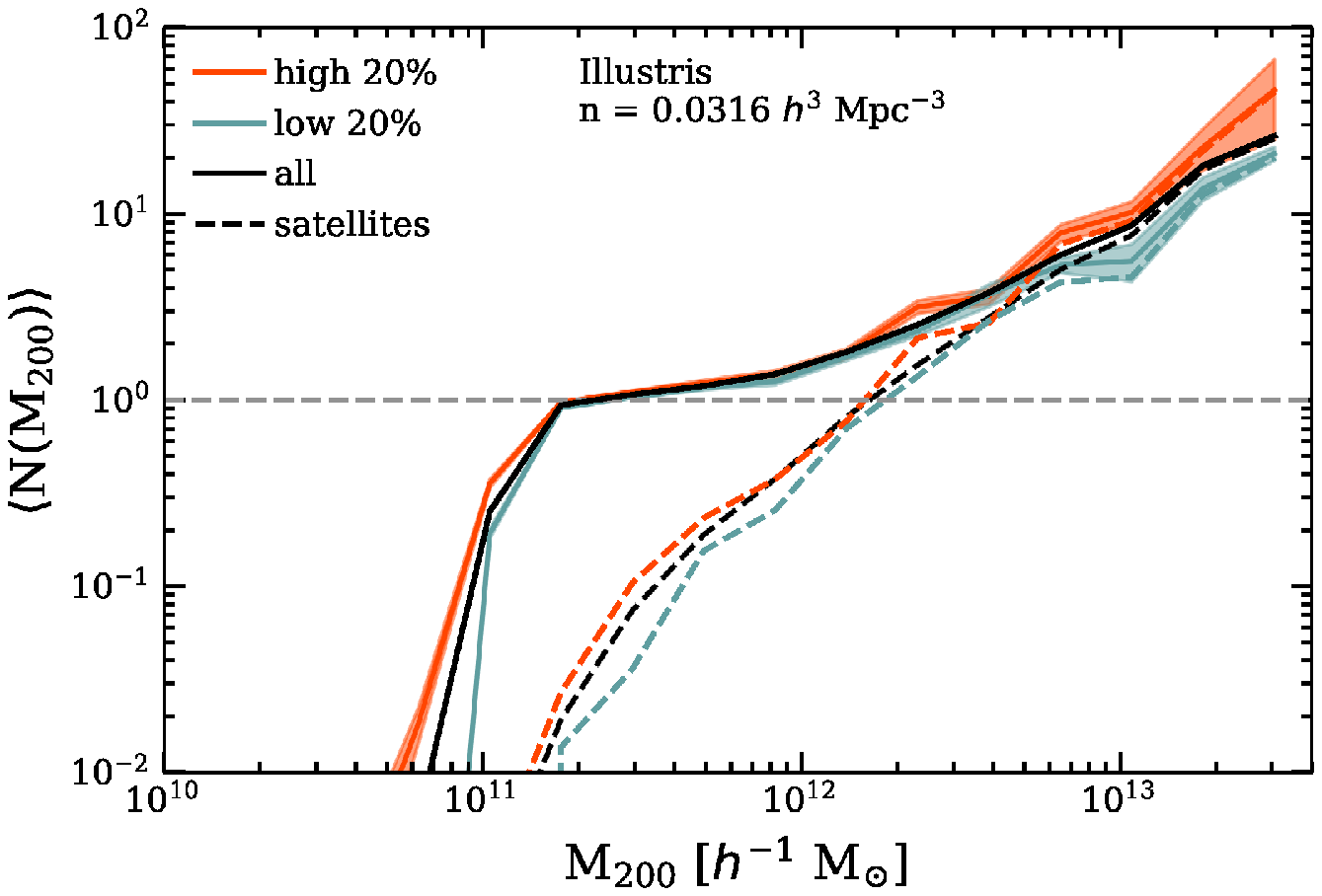}
 \includegraphics[width=0.45\textwidth]{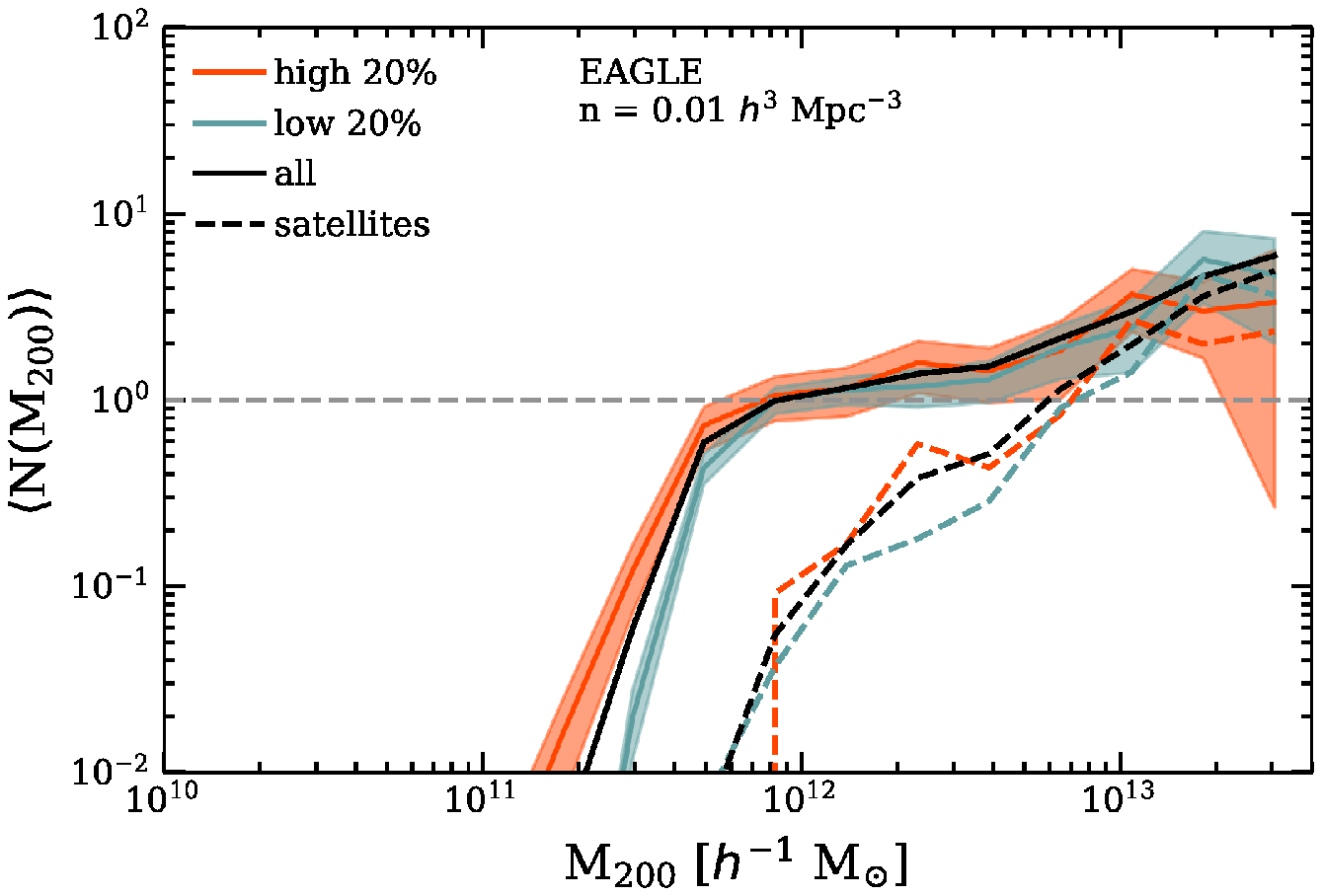}
 \includegraphics[width=0.45\textwidth]{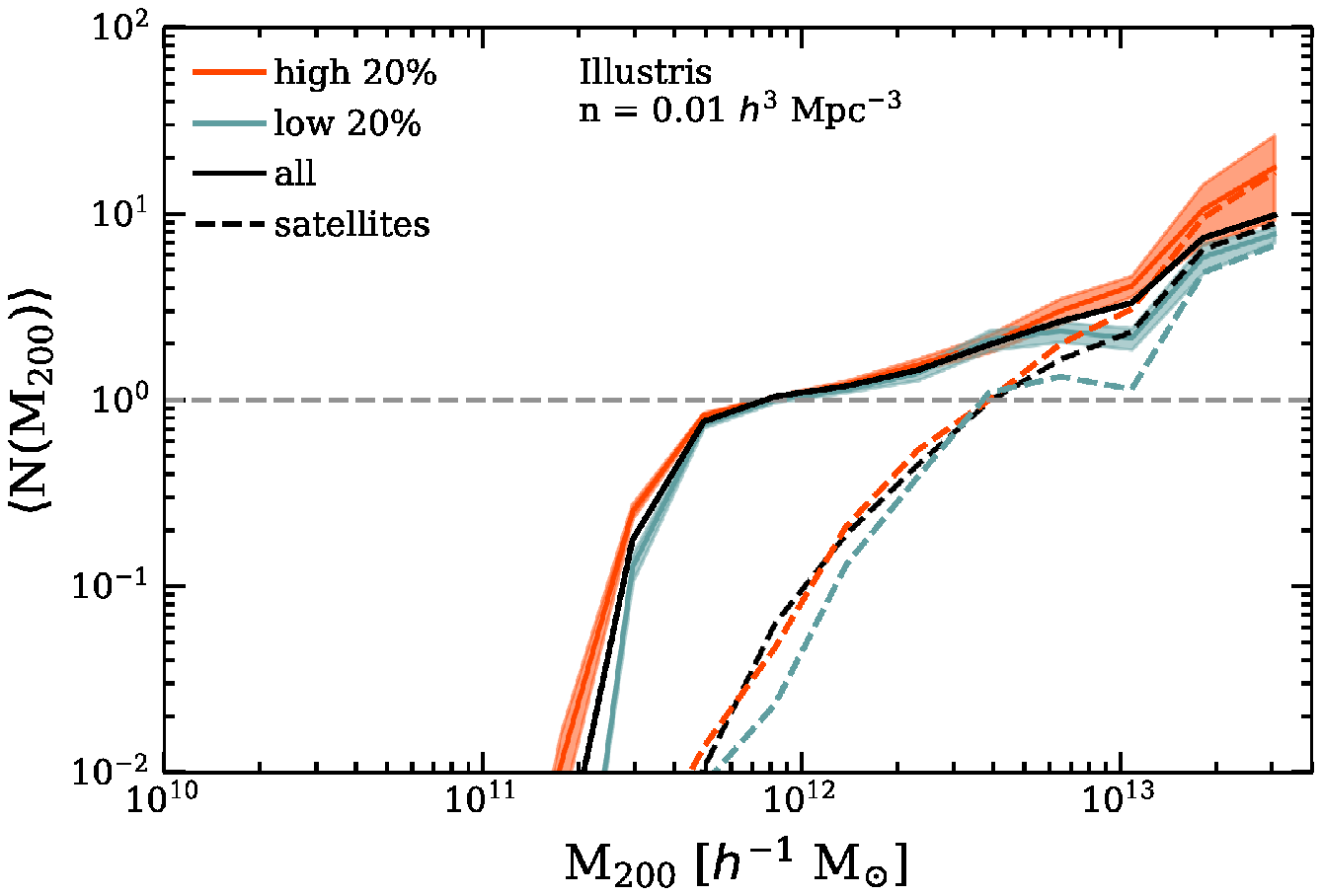}
 \includegraphics[width=0.45\textwidth]{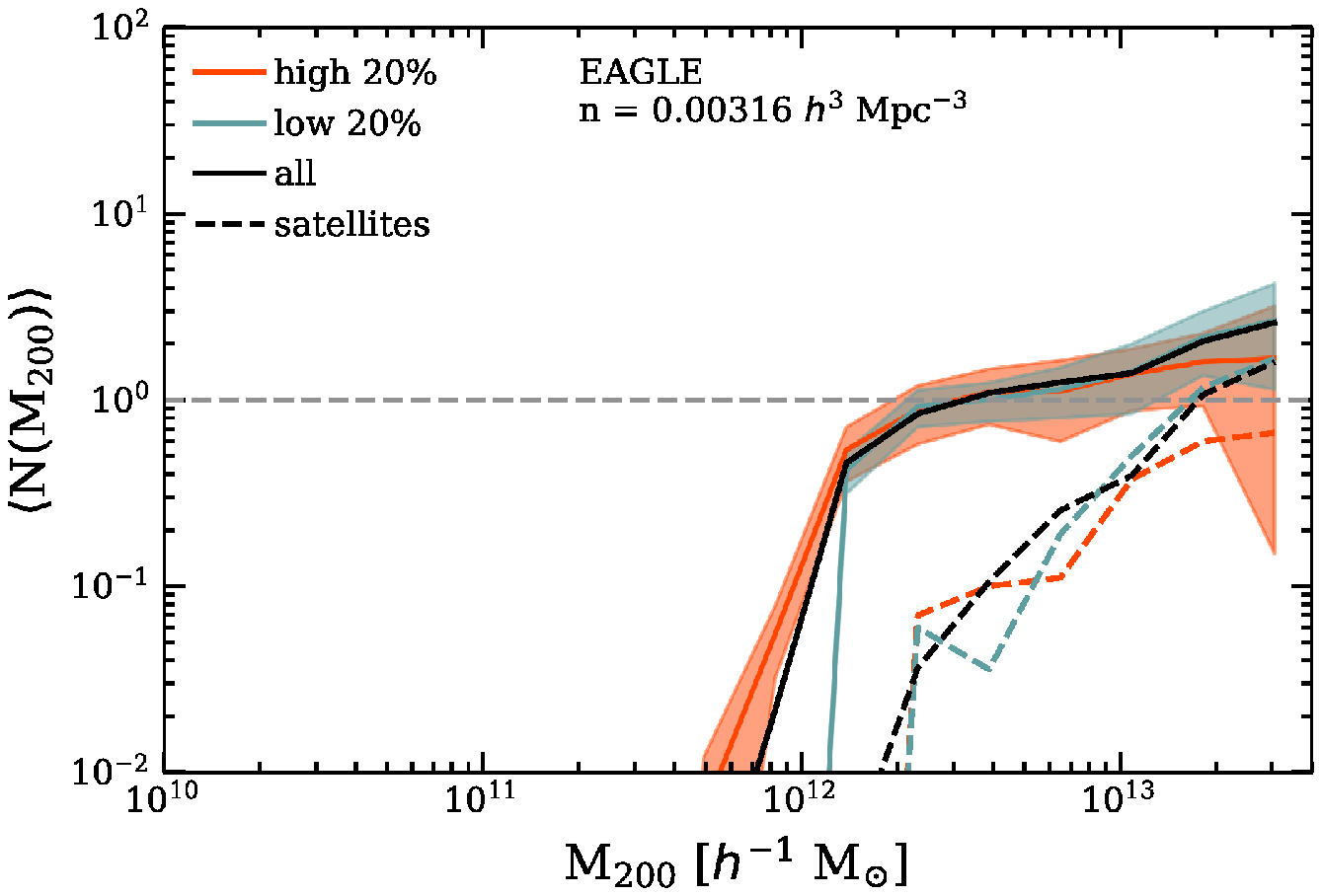}
 \includegraphics[width=0.45\textwidth]{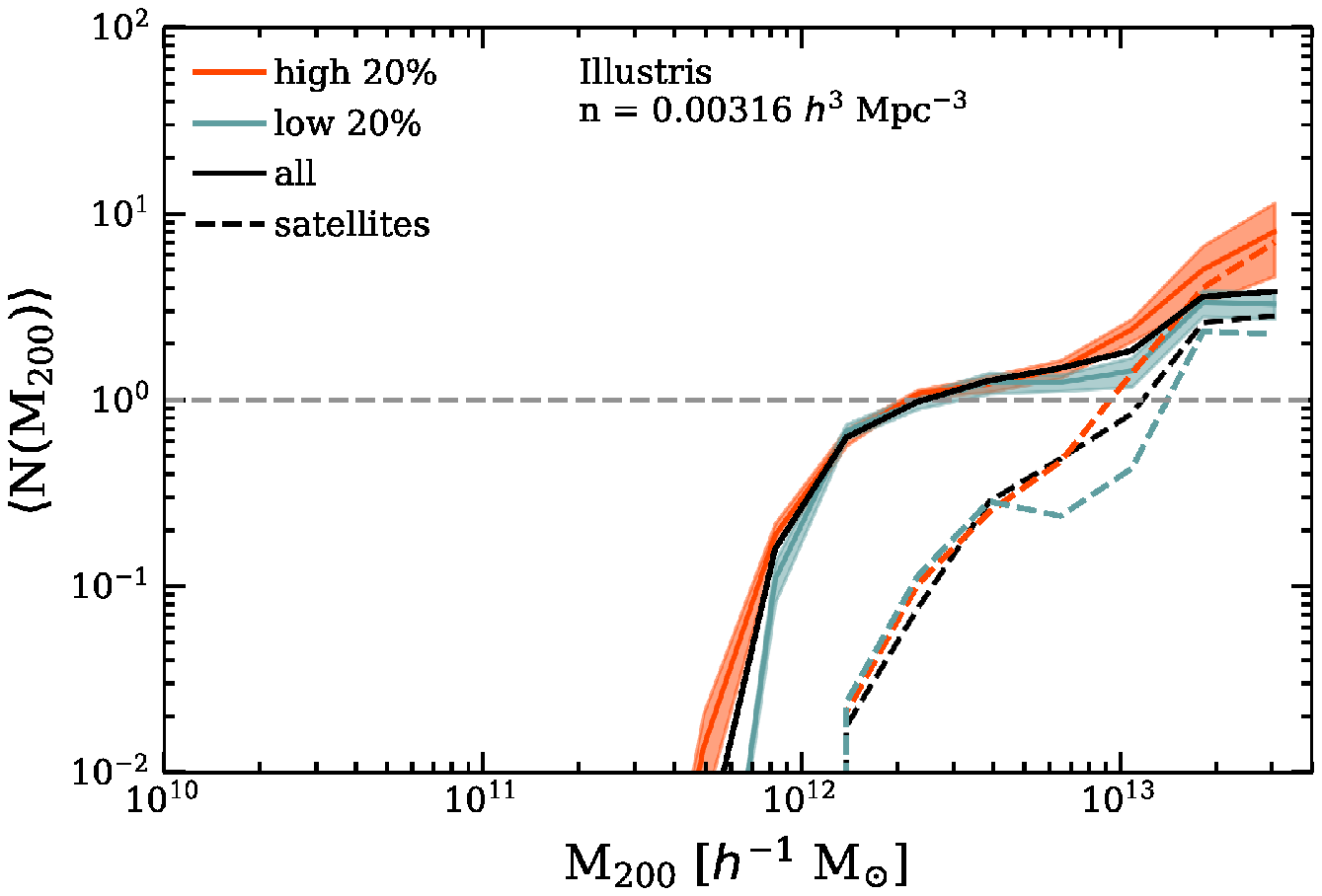}
  \caption{Halo occupation functions dependence on large-scale environment in
  \eagle\ (\textit{left panel}) and Illustris (\textit{right panel})
  for three number densities selected (high to low from top to bottom panel).
  Black solid lines represent the HOD of all galaxies in each sample. The red solid lines 
  represent the HOD of the galaxies in the 20\% of haloes in the densest regions, while the blue solid lines represent the HOD for 
  the galaxies in the 20\% of haloes in the least dense regions (following the definition described in Sec.~\ref{sec:definitions}).
  Error bars for the red and blue lines are computed via jack-knife resampling using 27 sub-volumes.
  Dashed lines represent the satellite occupancy functions for each sample.
  }
  \label{fig:HODEnvironment}
 \end{figure*} 
 
In Fig.~\ref{fig:HODEnvironment} we present the HOD as a function of environment 
for our three number density cuts n$ = (0.0316,0.01,0.00316) h^3$~Mpc$^{-3}$.
We obtain jack-knife error bars for the HOD of the galaxy samples for the most/least dense regions by using 
27 sub-volumes of the full simulated boxes. 
We limit the halo mass range below $M_{200} \sim 3.5\times10^{13}$~$h^{-1}~{\rm M}_{\odot}$ in order to have at least ten
dark matter haloes per mass bin.

In both simulations at low halo masses ($M_{200} \lesssim 10^{12} h^{-1} {\rm M}_{\odot}$) 
we find clear signatures of occupancy variation in the turnover of the central HOD, where the 
high density environments more likely to host a central galaxy than those in low dense environments. 
The trend is more significant in \eagle\ than Illustris for all number densities.
The occupancy variation reduces for the massive haloes, where haloes in most and least dense environments show a similar mean number of galaxies.  

In Fig.~\ref{fig:HODEnvironment}  we also show the satellite HOD (dashed lines) for the complete sample and for the haloes in
the most and least dense environments.
For the number density of  n$ = 0.0316~h^{3}$Mpc$^{-3}$, we find that the haloes in the densest regions
from \eagle\ and Illustris shift toward lower halo mass, like the central occupancy variation.
This may be a consequence of the densest environments favouring halo interactions and mergers.  
However, this trend is not clear for the other number densities studied, possibly due to the limitations of the simulated volumes.  
The error bars in Illustris are much narrower than \eagle\/. This might be due to haloes in the densest
environments being more uniformly distributed in Illustris (from an inspection by eye of Fig.~\ref{fig:SpaceDistrib}).
Finally, both simulations suggest that besides their mass, the halo occupation depends on the large-scale environment,
in agreement with recent findings \citep{McEwen2018,Zehavi2017} but in contrast to some earlier studies with hydrodynamical simulations
\citep{Berlind2003,Mehta2014}.

\subsection{The HOD dependence on halo formation time}\label{sec:HODTime}
 
  \begin{figure*}
 \centering
 \includegraphics[width=0.45\textwidth]{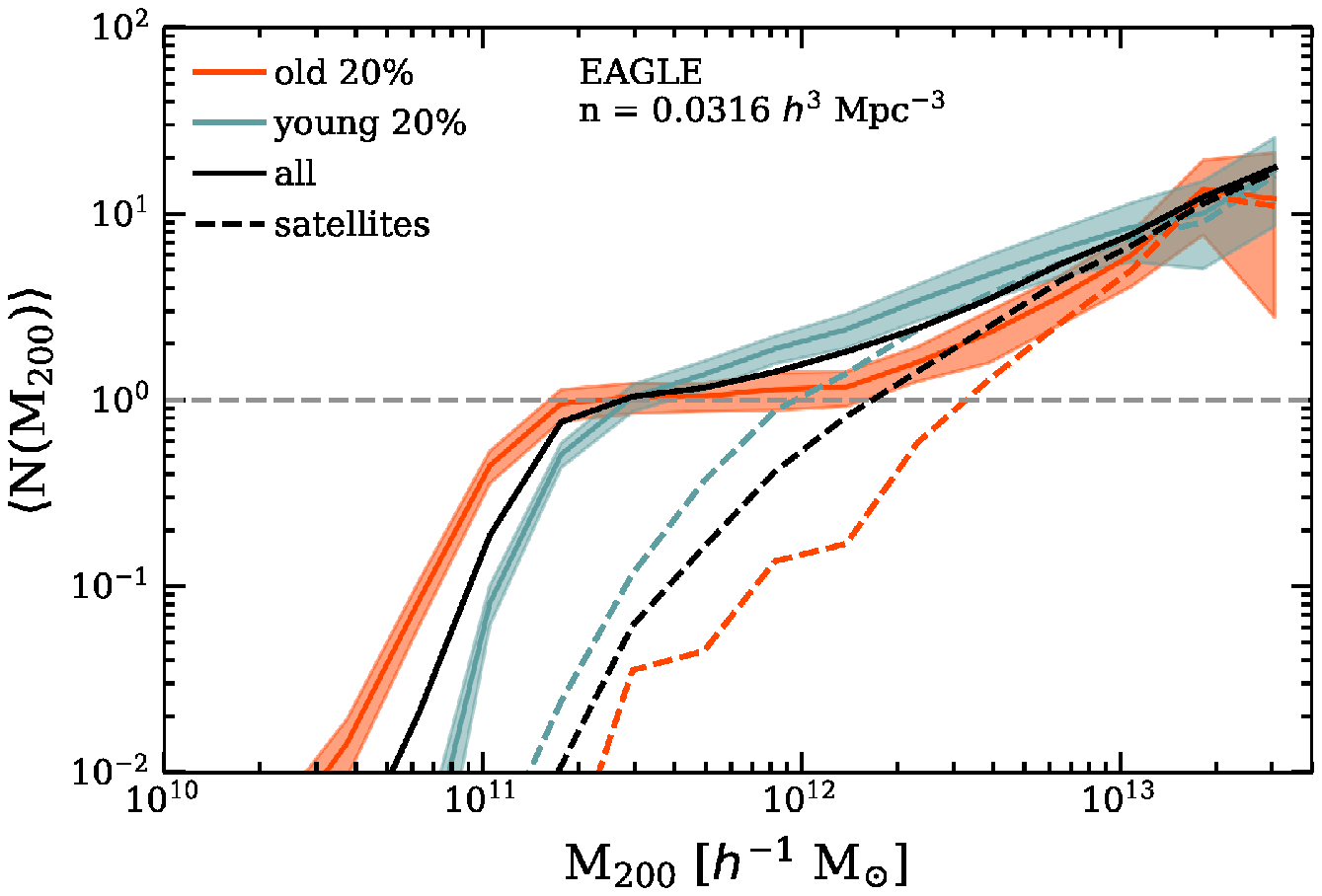}
 \includegraphics[width=0.45\textwidth]{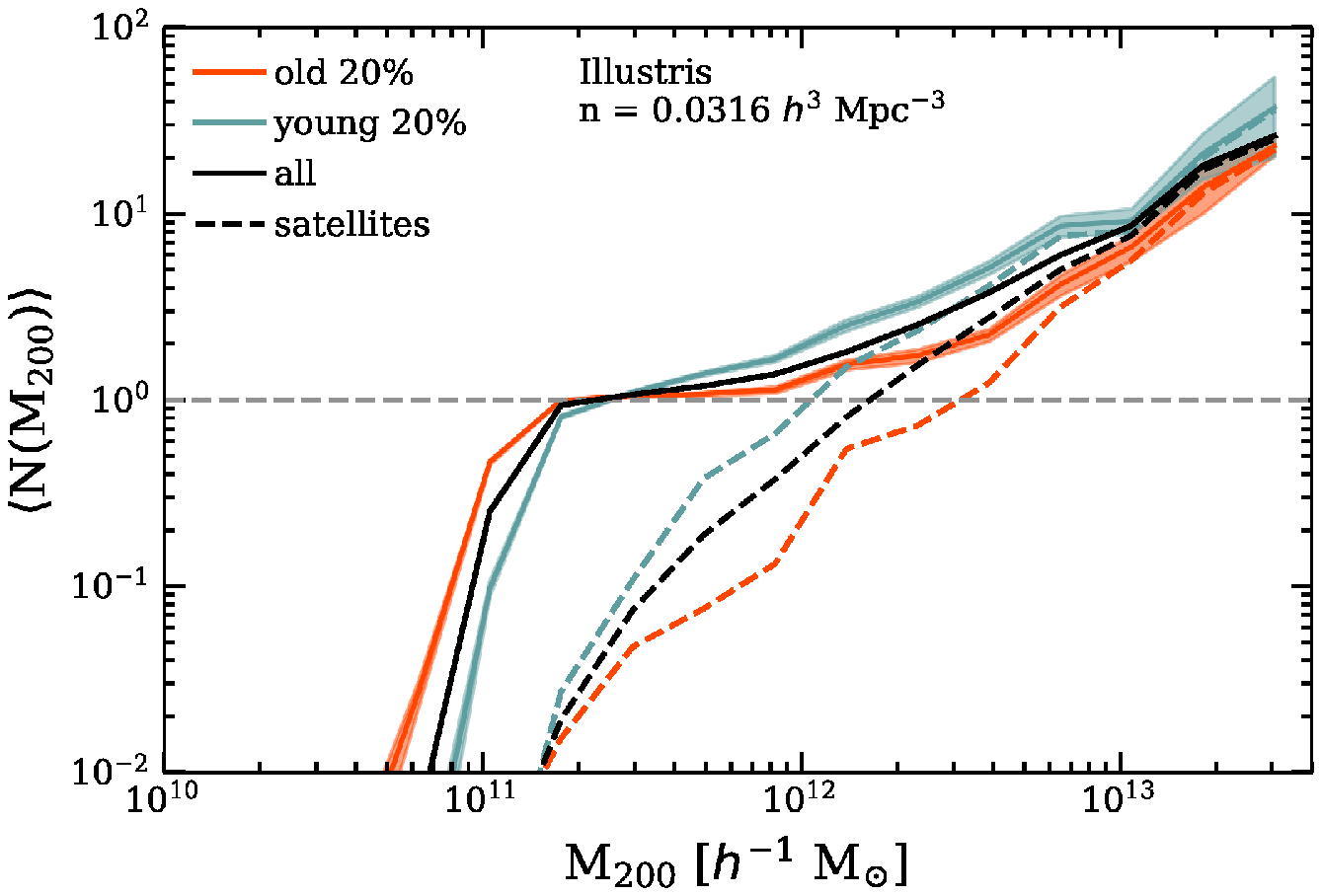}
 \includegraphics[width=0.45\textwidth]{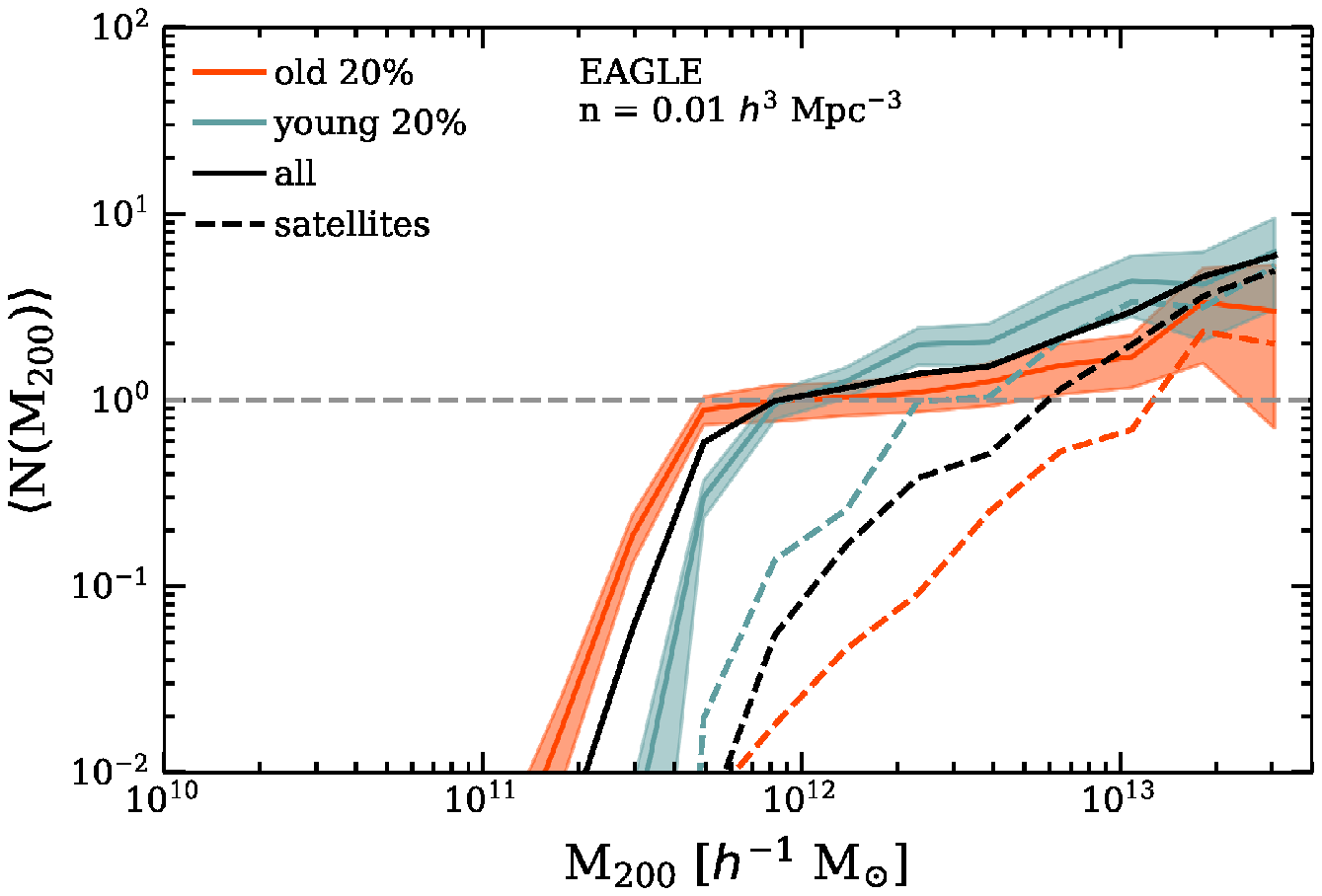}
 \includegraphics[width=0.45\textwidth]{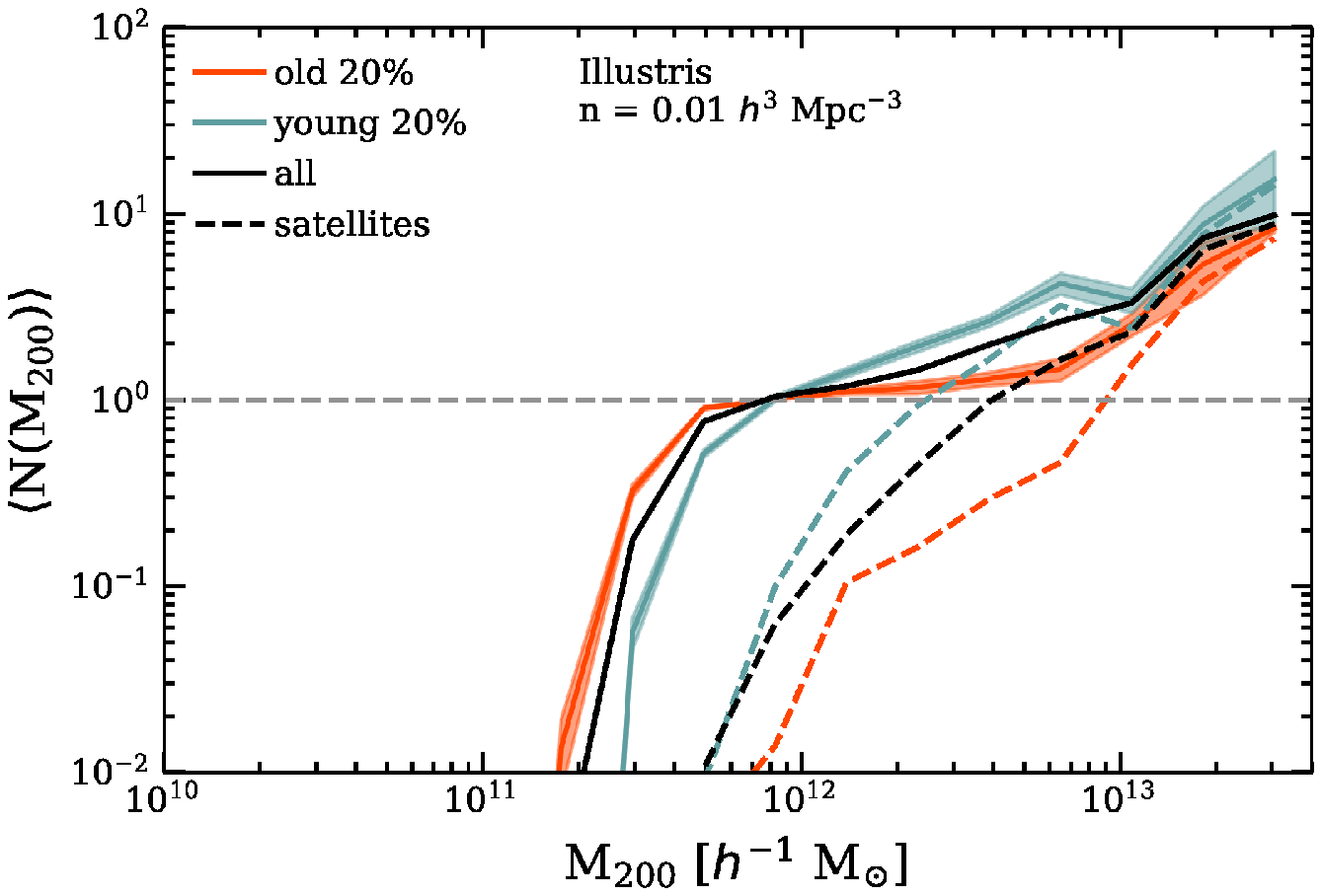}
 \includegraphics[width=0.45\textwidth]{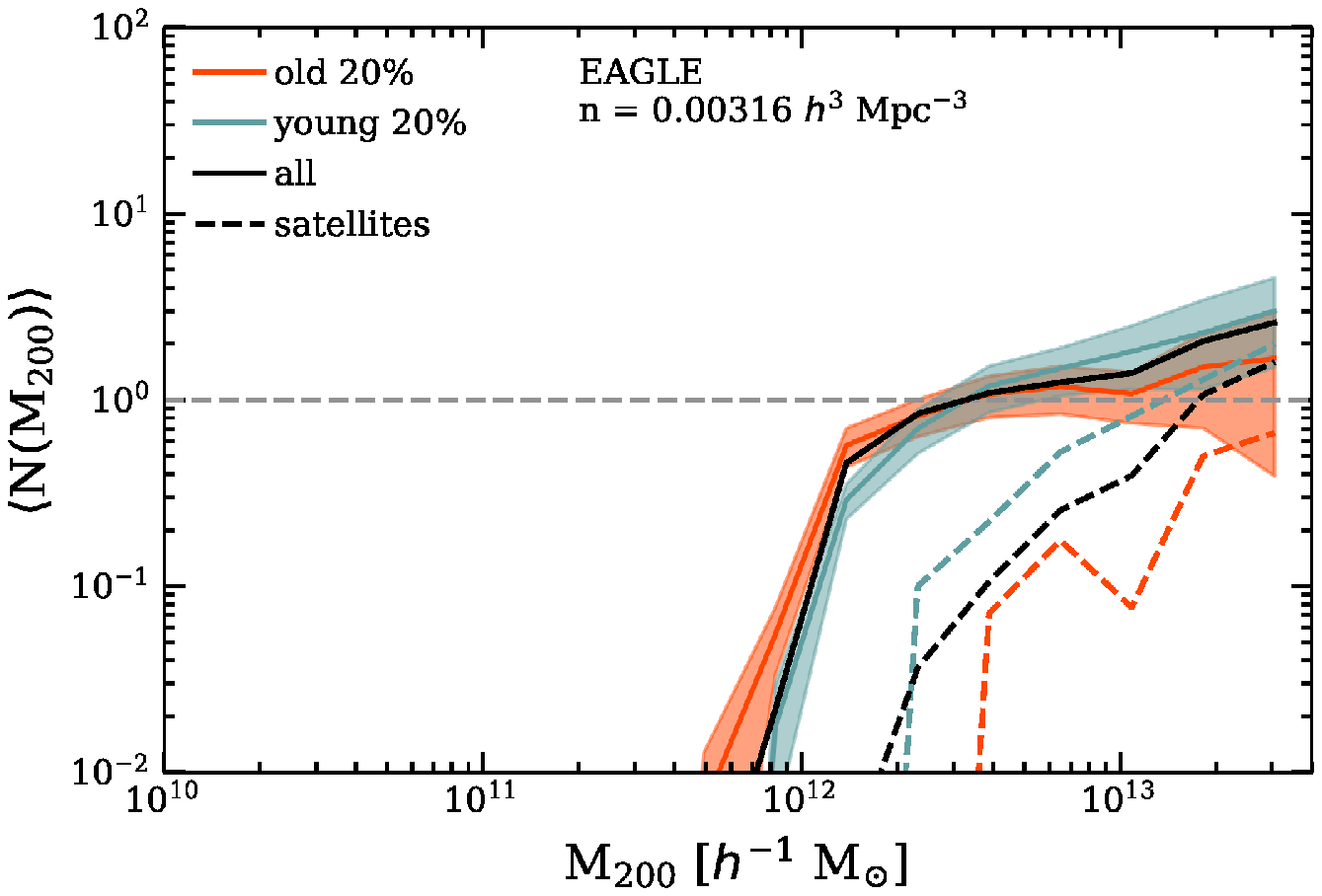}
 \includegraphics[width=0.45\textwidth]{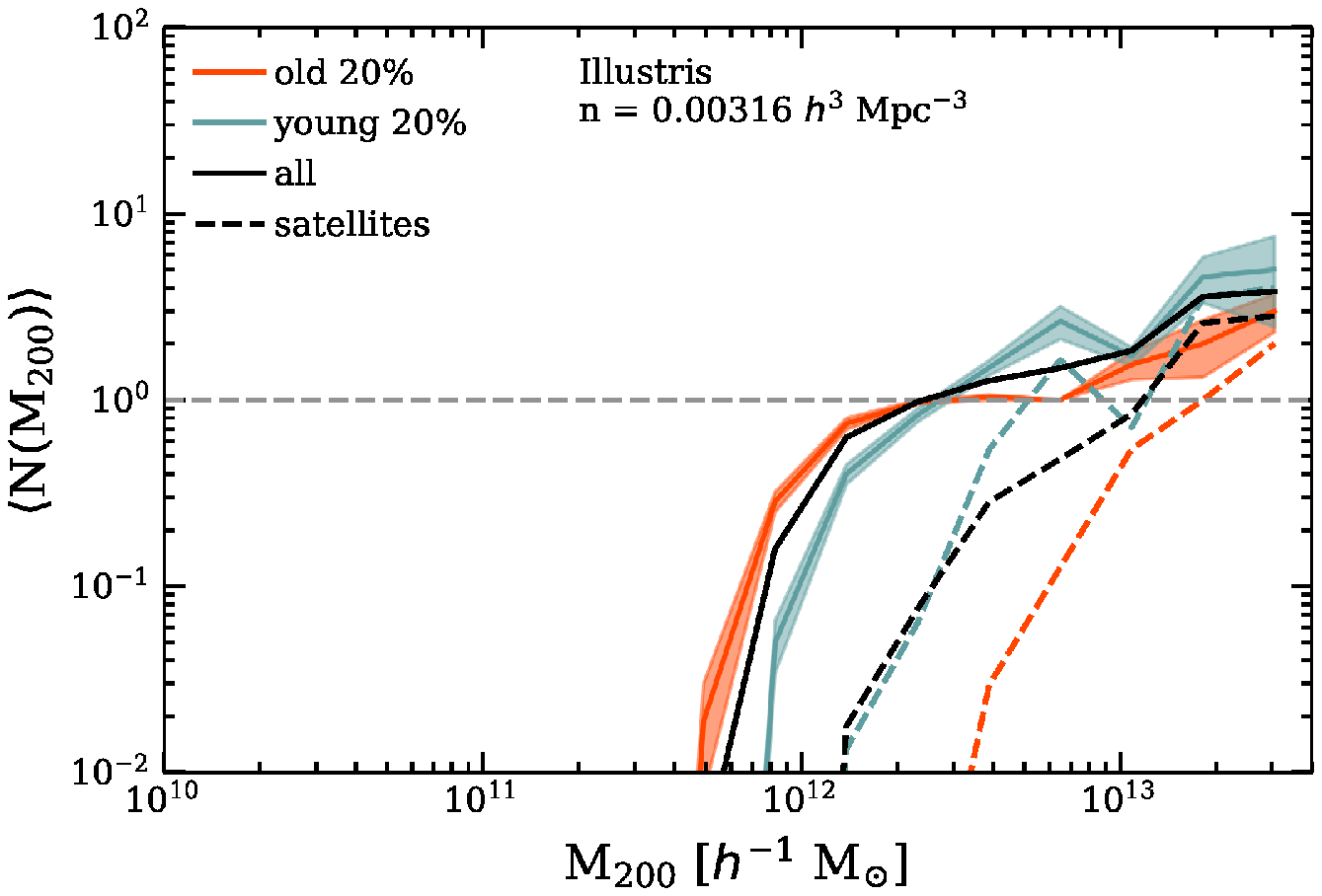}
  \caption{Halo occupation functions variation with halo formation time  
  for \eagle\ (\textit{left panel}) and Illustris (\textit{right panel})
  for the three number densities selected n~$= 0.0316, 0.01, 0.00316$~$h^{3} {\rm Mpc}^{-3}$ 
  (\textit{top}, \textit{middle} and \textit{bottom panel}, respectively).
  Black solid lines represent the HOD of the galaxies in each sample. 
  The red solid lines show the HOD for the galaxies in the 20\% early-formed (old) haloes, while blue solid lines show the HOD for 
  the galaxies in the 20\% late-formed (young) haloes, following the definition described in Sec.~\ref{sec:definitions}.
  We include error bars for the red and blue lines computed through jack-knife resampling by using 27 slices of the full simulated volume.
  Dashed lines show the satellite occupancy functions for each sample.}
  \label{fig:HODFormationTime}
 \end{figure*} 

Fig.~\ref{fig:HODFormationTime} presents the HOD for \eagle\ and Illustris for the complete sample of each number density
and for the galaxy populations in the 20\%  early-formed and late-formed haloes.
We include error bars for the galaxy samples in the early-formed and late-formed haloes (red and blue lines, respectively) 
computed using jack-knife resampling using 27 sub-volumes of each full simulated box.

Both simulations show that at low masses, the oldest haloes are more likely to host a galaxy than the youngest haloes.
This effect is mostly apparent in the ``turnover'' of the centrals occupation function.
We find that this result is clearer for \eagle\ than Illustris.
This trend reverses for high mass haloes (above $\sim 10^{12} h^{-1} {\rm M}_{\odot}$), where 
young haloes have on average a larger number of galaxies than old haloes.
This is explained by the satellites occupation shown as well in Fig.~\ref{fig:HODFormationTime} as dashed lines.
We see that the oldest haloes tend to host fewer satellite galaxies, likely
since they had more time to merge with the central galaxies.
This trend is found in both simulations and more clear for the intermediate and highest number densities.
Our findings are also in agreement with those from subhalo occupation split by halo formation time \citep[][see Appendix~\ref{apend:NsatNsub} for further discussions regarding this aspect]{vandenBosch2005,Mao2015,Jiang2017}.
Also \citet{Garaldi2018} using zoom-in hydrodynamical simulations find that the fraction
of mass in substructures is substantially larger in accreting (late forming) haloes.

We find in both \eagle\ and Illustris, that the variations in the HOD are stronger when split by formation time compared
to a split on environment.
This is again in agreement with \citet{Zehavi2017} who found similar results in two semi-analytic models. Hence, 
together with the previous findings, our results support the notion that the formation time is a more fundamental
property to describe the occupancy variation.

\section{The stellar mass - halo mass relation}\label{sec:SMHM}

 \begin{figure*}
  \centering
  \includegraphics[width=0.45\textwidth]{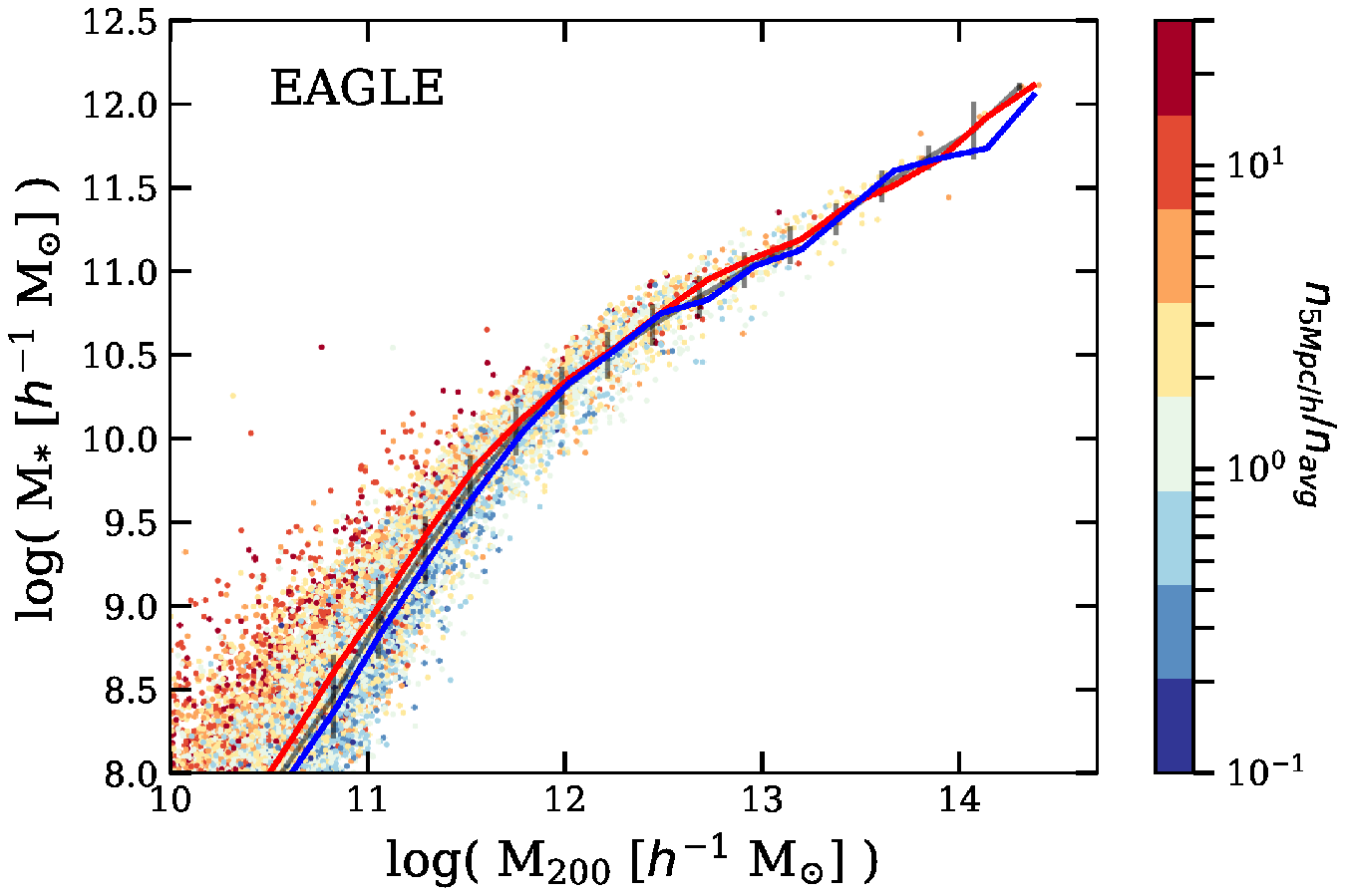}
  \includegraphics[width=0.45\textwidth]{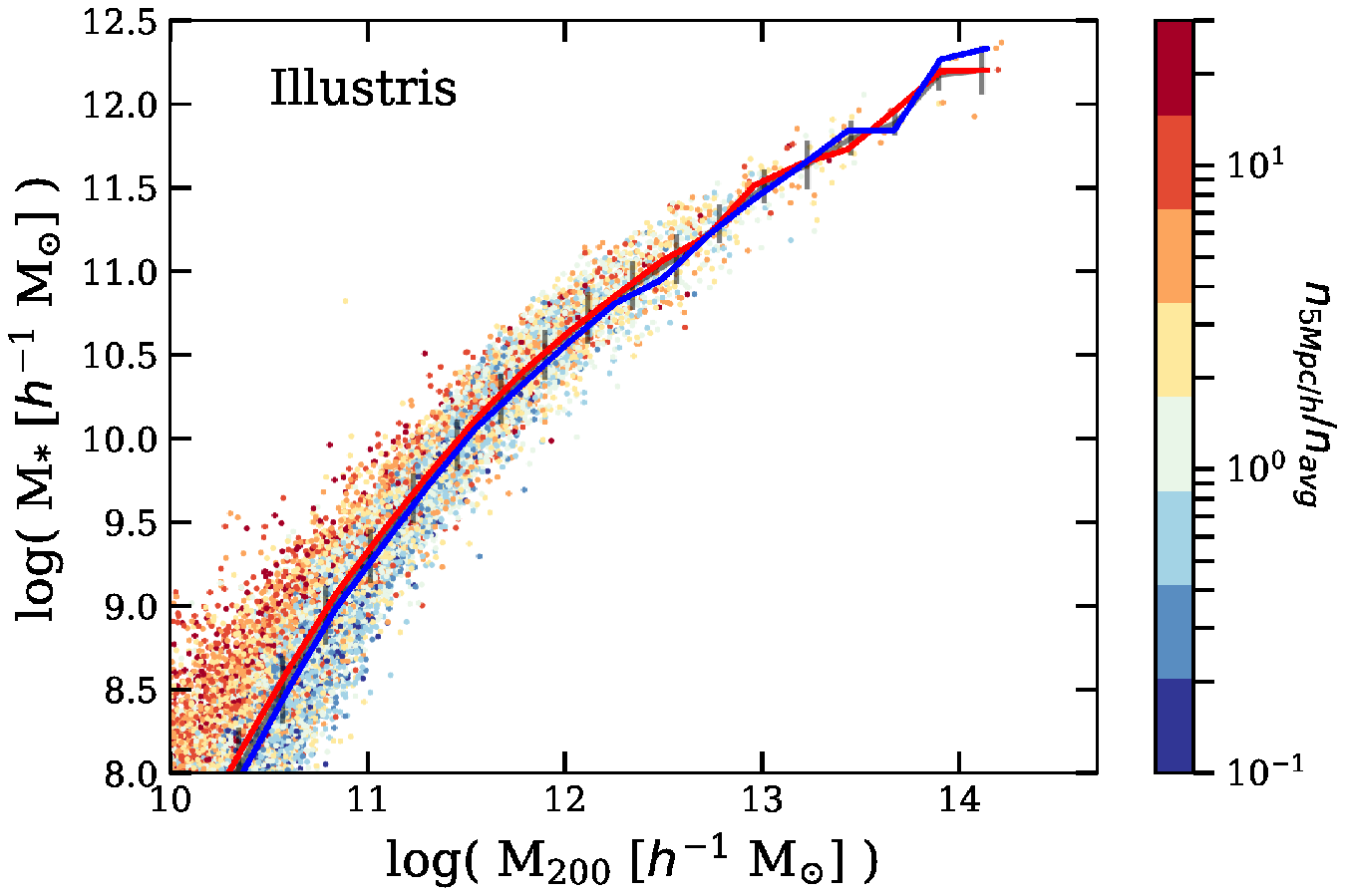}
  \caption{Stellar mass -- halo mass relation for the central galaxies in \eagle\ (\textit{left panel}) and Illustris simulation
  (\textit{right panel}), colour coded as a function of the environment of their haloes (n$_{\rm 5Mpc/h}$/n$_{avg}$).
  Black lines represent the median value of the distribution with the error bars representing
  20\% and 80\% percentiles. Red and blue lines
  show the median values for the galaxy samples within the most and least dense environments, respectively. 
  }
  \label{fig:MsMh-Env}
\end{figure*}

 \begin{figure*}
 \centering
 \includegraphics[width=0.45\textwidth]{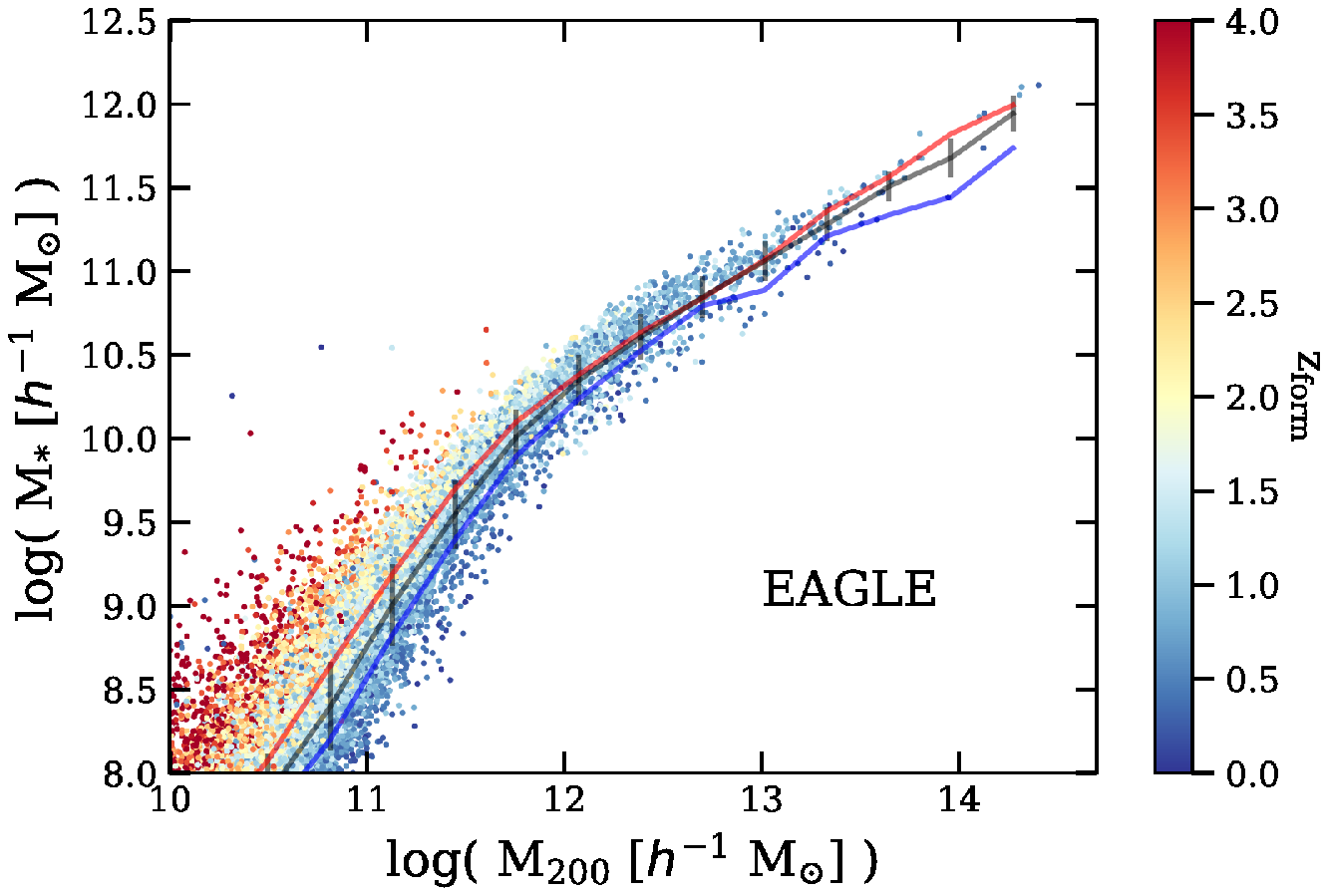}
 \includegraphics[width=0.45\textwidth]{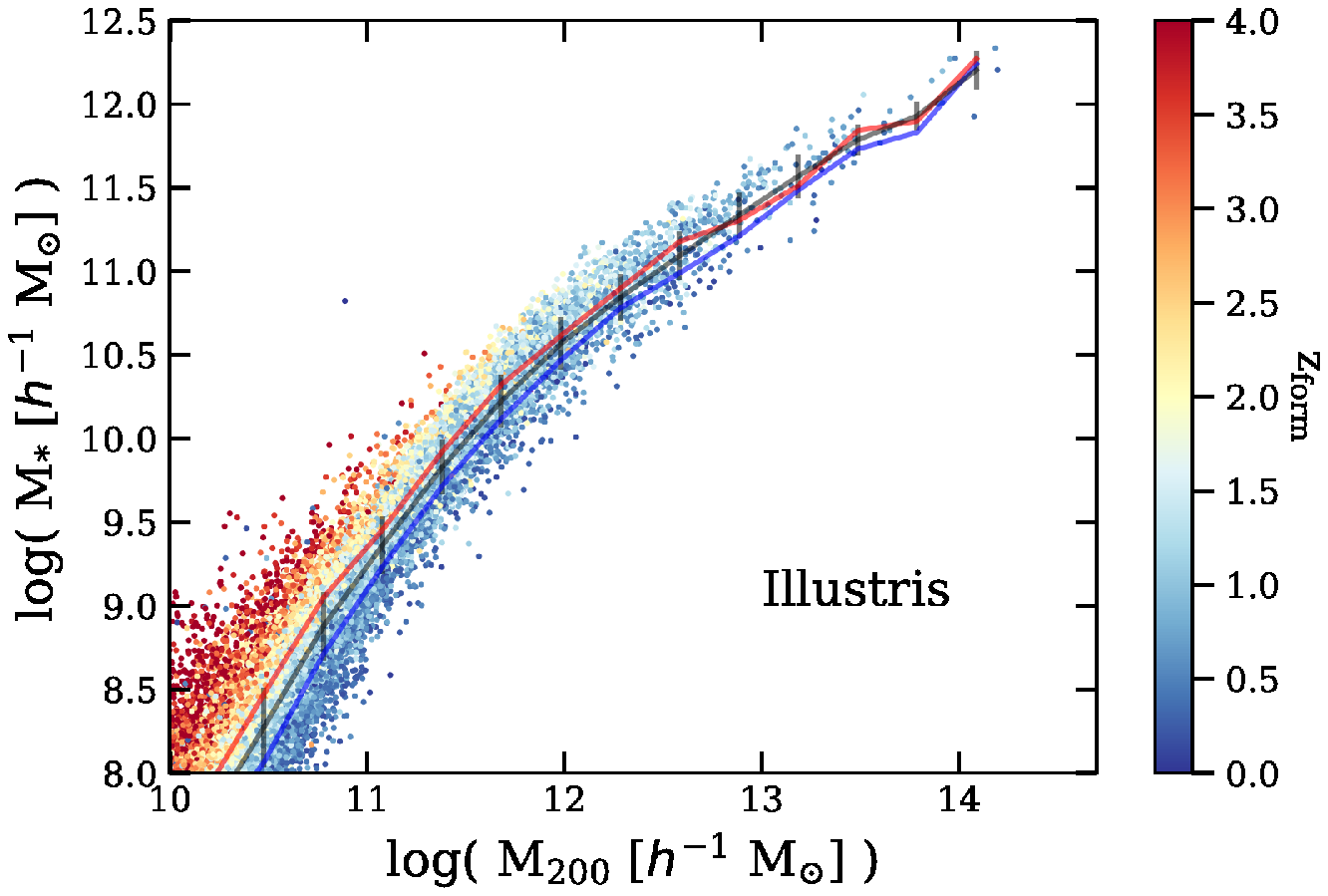}
  \caption{
  Stellar mass -- halo mass relation for the central galaxies in \eagle\ (\textit{left panel}) and Illustris simulation
  (\textit{right panel}), colour coded as a function of the formation time of their haloes ($z_{\rm form}$).
  Black lines represent the median value of the distribution with the error bars representing 20\% and 80\% percentiles.
  Red and blue lines show the median values for the galaxy samples within the early-formed and late-formed haloes, respectively.
 }
 \label{fig:MsMh-FormTime}
 \end{figure*}

The trends we find in the halo occupation with large-scale environment and formation time of the haloes can also be discussed
in the context of the stellar mass -- halo mass relation (SMHM) for central galaxies.
Following \citet{Zehavi2017}, we show in Fig.~\ref{fig:MsMh-Env} the SMHM dependence on environment
for \eagle\ (\textit{left}) and Illustris (\textit{right}).
The black thin line shows the median value of the distribution, with 
error bars representing the 20\% and 80\% percentiles exhibiting, as expected, the stellar mass increase with halo mass.

The SMHM in both simulations shows a change in the slope at a halo mass of 
$M_{\rm 200} \sim 10^{12} h^{-1} {\rm M}_{\odot}$, as a consequence of the contribution from the different 
feedback mechanisms (supernovae and AGN feedback) and merger mechanisms to the stellar mass of galaxies \citep[see, e.g.,][]{Matthee2017}. 
This change in the slope is clearer for \eagle\ than Illustris.
The dispersion of the SMHM is dependent on halo mass, where
dark matter haloes below to  $\sim 10^{12} h^{-1} {\rm M}_{\odot}$ present a larger range of stellar masses for the central galaxies.
As discussed in \citet{Matthee2017}, the dispersion at a fixed halo mass might be due to differences
in the halo concentration which in turn is related with their formation times. 

The central galaxies in Fig.~\ref{fig:MsMh-Env} are colour coded by their large-scale environment.
We see a clear dependence on environment, where for fixed halo mass, more massive central galaxies
reside preferentially in the denser environments.
We also show the median of the distribution for the central galaxies in the 
most and the least dense environments (red and blue lines, respectively).
Our results indicate that for both simulations, central galaxies in haloes below $\sim 10^{12} h^{-1} {\rm M}_{\odot}$
in the most dense environments are more massive than those in the least dense environments.
This is directly related to the results of the halo occupancy, where dark matter haloes in the densest environments 
are more likely to host central galaxies above a given stellar mass threshold
than those dark matter haloes in the least dense environments.

This SMHM trend seems to be more significant and with less scatter
for the hydrodynamical simulations studied in this work than in the semi-analytic models
studied in \citet{Zehavi2017}. 
This may be related to the apparent stronger occupancy variation signal in the hydrodynamical simulations compared to
the semi-analytic models. However, we cannot make a detailed comparison here due to the different environment definitions
used in \citet{Zehavi2017}.

The SMHM dependence on halo formation time for central galaxies in \eagle\ and Illustris is shown in Fig.~\ref{fig:MsMh-FormTime}.
Black line shows the median value of the distribution with the 20\% and 80\% percentiles, 
while red and blue lines are the median of the distribution for those galaxies within the early-formed and late-formed haloes, respectively.
The colour coding is done now by halo age.
Our results show a wide range of formation times for the dark matter haloes below to $\sim 10^{12} h^{-1} {\rm M}_{\odot}$,
while massive haloes are in general late-formed, as also shown in Fig.~\ref{fig:DefEnv}.
In both simulations, for the entire halo mass range studied, we find at fixed halo mass that 
central galaxies in early-formed haloes are more massive than those in late-formed haloes.
This indicates that the halo formation time affects the occupancy of the haloes, in addition to halo mass.

Our findings thus show that, at fixed halo mass, there is a strong correlation between the 
age of the halo and the stellar mass of the central galaxy, as seen by the separation of the blue/red lines in Fig.~\ref{fig:MsMh-FormTime}.
It arises since central galaxies in early-formed haloes have more time to accrete mass and form more
stars and become more massive.
This trend is slightly weaker when splitting the halo population by large-scale environment.
As discussed in \citet{Zehavi2017}, these results show that the stellar mass of the central galaxy
depend on other properties besides the halo mass, directly giving rise to 
the central galaxy occupancy variation with halo formation time or environment.


\section{Summary and Conclusions}
\label{sec:conclusions}

In this work we study the signals of the occupancy variation of the HOD in the state-of-the-art
hydrodynamical cosmological simulations \eagle\ and Illustris.
These occupancy variations coupled with halo assembly bias are what give rise to galaxy assembly bias.
This work represents an extension of the recent study of \citet{Zehavi2017}
with two semi-analytic models.

For each simulation, we compare the galaxy population within the dark matter haloes more massive than $10^{10}$~$h^{-1}$M$_{\odot}$
selected by large-scale environment and halo formation time. 
We define the large-scale environment of each halo by counting the number of subhaloes within a sphere of 5~$h^{-1}$Mpc radius, while the
halo formation time is estimated as the redshift for which the halo has assembled half of its present-day mass.
In order to find the differences in the extreme cases, we select the 20\% of the haloes in the 
most/least dense environment and the 20\% latest/earliest formed haloes, in bins of halo mass.
We analyse three fixed number density samples ranked by stellar mass corresponding to 
n$ = 0.0316, 0.01, 0.00316$~$h^{3}{\rm Mpc}^{-3}$.

The comparison of the full HOD from \eagle\ and Illustris at different number densities show that the
mean occupation of haloes in Illustris is shifted towards lower halo masses relative to \eagle\
(see Fig.~\ref{fig:HOD-ND}). 
This can be explained by the differences in the halo mass functions of the
simulations. 

We find that the mean occupation of low mass haloes ($\lesssim 10^{11}$--$10^{12}$~$h^{-1}$M$_{\odot}$)
depends on the large-scale environment in both simulations, and this result is present for the three galaxy number densities selected.
Thus, haloes in the densest regions are more likely to host a central galaxy than those in the least 
dense environments (Fig.~\ref{fig:HODEnvironment}). 
Moreover, the satellite occupation of the haloes shows a slight dependence on the large-scale environment, where
the haloes in the densest regions have a mean occupation higher than those in least dense regions.

Examining the stellar mass -- halo mass relation for the central galaxies, we find that, at a fixed dark matter 
halo mass below to $10^{12}$~$h^{-1}$M$_{\odot}$, the central galaxies in the densest regions
are more massive than those in the least dense environments (Fig.~\ref{fig:MsMh-Env}).
These results are present in both simulations, although \eagle\ shows a stronger difference between the populations of
haloes in least/most dense regions. 
Furthermore, our findings are in general agreement with the results presented in \citet{Zehavi2017} for semi-analytic
galaxy formation models, and with those from \citet{McEwen2018} using age-matching mock catalogues.
This is in contrast with earlier analyses \citep{Berlind2003,Mehta2014} using different hydrodynamical simulations,
which do not report the observed trends seen in this work.
These differences perhaps have to do with recent improvements in the stellar and AGN feedback models, 
and we plan to further investigate this in future work.

When we split the dark matter haloes by their formation time, we find a more significant difference in
the halo occupation.
Both simulations show that at low mass the early-formed (old) haloes are more likely to have a central galaxy than
the late-formed (young) haloes, for the three cumulative number densities (see Fig.~\ref{fig:HODFormationTime}).
This trend reverses at high masses due to the contribution 
of the satellite galaxies. Hence, we find that late-formed haloes host more satellite galaxies than early-formed haloes.
These results can be explained by the central galaxies having more time to assemble in the early-formed haloes. And in turn, the
satellite galaxies have more time to merge with the central galaxy in early-formed massive haloes. 

Finally, we study the SMHM relation in terms of halo formation time, finding that, at fixed mass,
the early-formed haloes host more massive central galaxies than the late-formed haloes. 
This result supports the idea that early-formed haloes have more time to form a massive central galaxy 
and gives rise to the measured occupancy variations.

In agreement with previous findings by \citet{Zehavi2017}, 
we find evidence of occupancy variation when splitting the haloes either by
their formation time or by the large-scale environment, for both hydrodynamical simulations.
Furthermore, we find that this signal is more significant when we split the halo population by age. 
These results provide robust predictions for occupancy variation from the latest state-of-the-art galaxy formation models.
It remains to be determined what is the extent of occupancy variation and galaxy assembly bias in the real Universe.

\section*{Acknowledgement}
We thank the anonymous referee for insightful comments.
We also thank Carlton Baugh and Nelson Padilla for useful discussions and comments.
MCA and SC acknowledge the Southern Astrophysics Network Proyecto Redes 150078 (Conicyt-Chile) for
partial financial support, and the organizers of the first workshop held in Santiago de Chile where this project began.
IZ acknowledges support by NSF grant AST-1612085.
SC acknowledge support from a STFC/Newton-CONICYT Fund award (ST/M007995/1DPI20140114),
Anillo ACT-1417 and the European Research Council through grant number ERC-StG/716151.
PN acknowledges the support of the Royal Society through the award of a University 
Research Fellowship, 
and the Science and Technology Facilities Council (ST/P000541/1). 
We acknowledge the Virgo Consortium for making their simulation data available. The \eagle\ simulations
were performed using the DiRAC-2 facility at Durham, managed by the ICC, and the PRACE facility
Curie based in France at TGCC, CEA, Bruy\`{e}resle-Ch\^{a}tel.

\bibliographystyle{mnras}
\bibliography{Artale-HOD}

\appendix

\section{The definition of density environment}\label{sec:app}

In \S~\ref{sec:definitions} we define the criteria to select the dark matter haloes in dense and underdense environments.
In summary, for each dark matter halo with $M_{200} > 10^{10} h^{-1} {\rm M}_{\odot}$,
we compute the number density within a sphere of fixed radius by counting the number
of subhaloes using periodic boundary conditions, and excluding those within the halo ($n_{R_{max}}$) and dividing by the volume of the sphere.
We count the subhaloes within the sphere but outside of the halo itself
since we want to measure the environment on large scales. 
This value is normalized by the mean number density of subhaloes in haloes of M$_{200}>10^{10}~h^{-1} {\rm M}_{\odot}$ ($n_{avg}$).
Thus, a ratio of one indicates that the halo resides in the average density environment.
In particular, we then select the 20\% of haloes in the most and least dense environments,
thus providing a way to quantify and compare the halo occupation for the two extreme cases.

\begin{figure}
 \centering
 \includegraphics[width=0.45\textwidth]{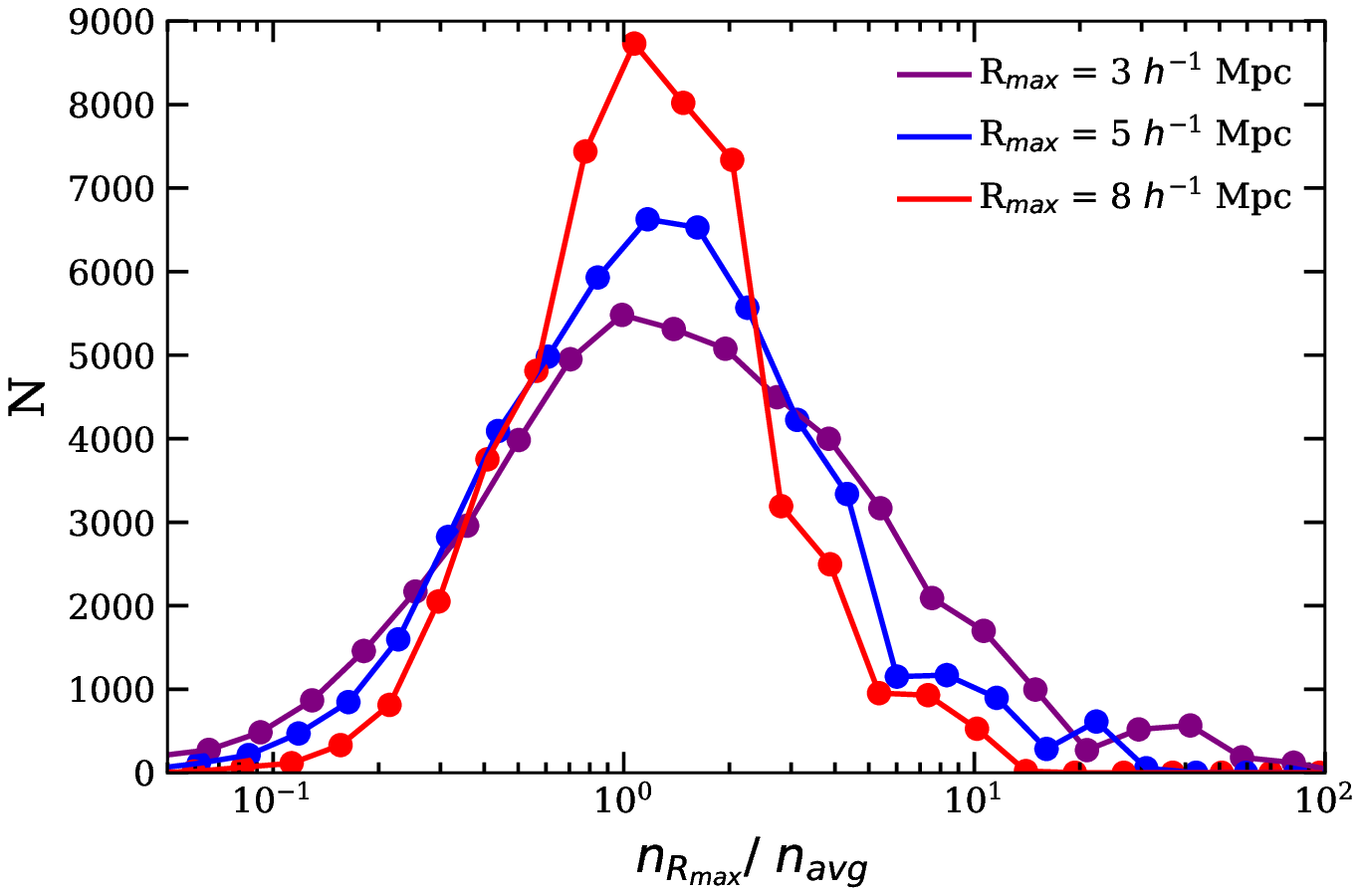}
 \caption{Distribution of the environmental densities of dark matter haloes with $M_{200} > 10^{10} h^{-1} M_{sun}$,
 in a sphere of $R_{\rm max} = 3, 5, 8 h^{-1} Mpc$ at $z = 0$ (purple, blue and red, respectively).
 We compute the number of subhaloes within the sphere of $R_{\rm max}$ ($n_{R_{max}}$), and normalize by the average number density  
 of subhaloes in dark matter haloes of $M_{200} > 10^{10} h^{-1} M_{sun}$ ($n_{avg}$).}
 \label{fig:HistoEnv}
\end{figure}

We check here the impact of the sphere radius used to calculate the environment.
In order to discuss how robust is the radius selected, in Fig.~\ref{fig:HistoEnv} we show
the distribution of the environment densities computed with three different thresholds in $R_{\rm max} = 3, 5, 8~h^{-1}$Mpc.
Our results show that the distributions obtained with the three different $R_{\rm max}$ are similar, 
finding only expected differences in the number N according to the radius, where the larger radius shows a slightly narrower distribution 
and more haloes within the average density environment. 

We also test the impact of counting the subhaloes within each halo (i.e., for each halo we count all the subhaloes in the sphere
irrespectively if they belong to the halo or not) finding that the distribution is not affected.
Since we want to probe the large-scale environment, we would like to use a relatively large radius, significantly larger
than the typical halo size. At the same time, that radius should not be too large, so as to maintain a 
reliable sampling of a large range of environments.
Hence, we select the radius to be $R_{\rm max} = 5~h^{-1}$Mpc as a compromise of 
these two considerations as well as taking into account 
the size limits of the simulation. 
This radius represents roughly seven per cent of the simulated box side for both \eagle\ and Illustris.

\section{Fitting the HOD predicted by EAGLE and Illustris}\label{apend:fitsHOD}

The HOD is commonly parametrized by making a distinction between central and satellite galaxies \citep[e.g.,][]{Zheng2005,Zehavi2011,Contreras2013}.
For central galaxies, the mean occupation function can be described by a step-like function with a cut-off profile, while satellites follow a power-law function with a smooth cut-off at small halo masses.
The most frequently used parametrization is the 5-parameter model introduced by \citet{Zheng2005}. The model includes the quantities 
$M_{\rm min}$ and $M_{1}$, where $M_{\rm min}$ refers to the
halo mass for which half of the haloes on average host a central galaxy (i.e., $\langle {\rm N}_{\rm cen}(M_{\rm min}) \rangle = 0.5$), 
while  $M_{1}$ is the mass at which the halo has on average one satellite galaxy 
(i.e., $\langle {\rm N}_{\rm sat}(M_1) \rangle = 1$).
The ratio of these quantities ($M_{1}$/$M_{\rm min}$) gives us information about how much larger the dark matter halo mass has to be
in order to host an additional satellite galaxy beyond the central one.
Hence a larger value of the ratio implies a wider range of halo masses hosting just a central galaxy \citep{Berlind2003}. 
Furthermore, it influences the shape of the correlation function and reflects the balance between accretion and
destruction of satellite galaxies \citep{Zentner2005}.

The halo occupation function is commonly modelled in the following form \citep[following][]{Contreras2017,Zehavi2017}. For central galaxies:

\begin{equation}
 \langle N_{\rm cen}(M_{\rm h})\rangle = \frac{1}{2}\left[ 1 + {\rm erf} \left( \frac{\log M_{\rm h} - \log M_{\rm min}}{\sigma_{\log M}}  \right) \right],
\label{Eq:Cen_HOD}
\end{equation}

\noindent where $\sigma_{\log M}$ reflects the scatter between the halo and stellar mass, and 
$ {\rm erf}(x)$ is the error function,
$ {\rm erf}(x) = \frac{2}{\sqrt{\pi}} \int_{0}^{x} e^{-t^2} {\rm d}t. $
For satellite galaxies, we use:

\begin{equation}
 \langle N_{\rm sat}(M_{\rm h})\rangle = \left( \frac{M_{\rm h}-M_{\rm cut}}{M^*_1}\right)^\alpha,
\label{Eq:Sat_HOD}
\end{equation}

\noindent where $\alpha$ is the slope of the power-law, $M_{\rm cut}$ is the satellite cut-off mass scale, and $M^*_1$
is the normalization with, $M_{1} = M^*_1 + M_{\rm cut}$.
Therefore, the total halo occupation function is given by the sum of these two terms:
\begin{equation}
 \langle N_{\rm gal}(M_{\rm h})\rangle =  \langle N_{\rm cen}(M_{\rm h})\rangle +  \langle N_{\rm sat}(M_{\rm h})\rangle.
\end{equation}

We note that the model we implement, originally proposed by \citet{Zheng2005}, allows to fit 
the central and satellite occupation functions independently.
In Tables~\ref{tab:M1MminEAGLE} and \ref{tab:M1MminIllustris} we present the values obtained for these five
parameters of the halo occupation function for \eagle\ and Illustris.
We fit the HOD for central and satellite galaxies separately, and assume equal weight to all measurements
in the halo mass range that fulfills the condition of $\langle N(M_{\rm h})\rangle = 10^{-1.0} - 10^{1.5}$.
The fits are computed using the HOD with halo mass bins of $\sim0.24$~dex, the same 
as shown before.
We derive the errors from jack-knife resampling, using 27 sub-volumes for each simulation.
We find that $M_{1}$ and $M_{\rm min}$ increase as the number density cut decreases in both simulations. This is 
expected since a lower number density corresponds to a higher stellar mass cut. This is also the case for $M_{\rm cut}$.
We note here that the halo masses $M_{\rm h}$ implemented to compute the HOD correspond to those from the hydrodynamical 
simulations and not to their DMO counterparts. Hence, although we expect a negligible difference when using DMO halo masses, it
is important to consider this aspect when comparing our findings with, for example, SHAM models \citep[see][]{Chaves-Montero2016}.

For the ratio $M_{1}$/$M_{\rm min}$ we find different trends and values. For Illustris the ratio decreases with the 
number density, in agreement with previous results from observations and semi-analytic models \citep[e.g.,][]{Guo2014,Contreras2017},
while for \eagle\ the trend reverses. 
These results indicate that massive central galaxies, which reside in the more massive haloes,
are more likely to be accompanied by a satellite galaxy in Illustris than in \eagle\ .
We are unsure what is the cause of the reversed trend of $M_{1}$/$M_{\rm min}$ in the simulations studied.
It is also important to note that slight changes in the parameters $M_{1}$ and $M_{\rm min}$ 
result in different trends for the ratio, which further indicates that this trend should be regarded with caution.

\begin{table*}
 \caption{Values of the HOD parameters obtained from fitting the mean occupation
 function for satellites and central galaxies of \eagle.
 The units of $M_{\rm cut}$, $M_1$ and $M_{\rm min}$ are in $h^{-1} {\rm M}_{\odot}$.
 We present our results for the three number densities selected.
 The errors are computed through jack-knife resampling. 
 }
 \label{tab:M1MminEAGLE}
\begin{tabular}{|l|cccccc|}
\hline
                       &\multicolumn{6}{c|}{\sc eagle}\\              
n [$h^{3}$~Mpc$^{-3}$]  & $\alpha$          &  $\log M_{\rm cut}$  & $\sigma_{\log M}$ & $\log(M_{1})$             &   $\log(M_{\rm min})$ &   $M_{1}/M_{\rm min}$ \\
\hline
 3.16$\times 10^{-2}$   &  0.92$\pm$0.03   &   11.49$\pm$ 0.05    &  0.202$\pm$0.006   &    12.21$\pm$0.01         &    11.149$\pm$0.005   &   11.51$\pm$0.43      \\
 1.00$\times 10^{-2}$   &  1.08$\pm$0.12   &   11.34$\pm$0.81     &  0.243$\pm$0.017   &    12.78$\pm$0.03         &    11.656$\pm$0.009   &   13.30$\pm$0.84      \\
 3.16$\times 10^{-3}$   &  1.17$\pm$0.18   &   11.90$\pm$1.82     &  0.318$\pm$0.031   &    13.32$\pm$0.04         &    12.166$\pm$0.015   &   14.25$\pm$1.17     \\
\hline
\end{tabular}
\end{table*}

 \begin{table*}
 \caption{Values of the HOD parameters obtained from fitting the mean occupation
 function for satellites and central galaxies of Illustris.
 The units of $M_{\rm cut}$, $M_1$ and $M_{\rm min}$ are in $h^{-1} {\rm M}_{\odot}$.
 We present our results for the three number densities selected.
 The errors are computed through jack-knife resampling. 
 }
 \label{tab:M1MminIllustris}
\begin{tabular}{|l|cccccc|}
\hline
                        &\multicolumn{6}{c|}{Illustris}\\              
n [$h^{3}$~Mpc$^{-3}$]  & $\alpha$          &  $\log M_{\rm cut}$  & $\sigma_{\log M}$ & $\log(M_{1})$             &   $\log(M_{\rm min})$ &   $M_{1}/M_{\rm min}$ \\
\hline
3.16$\times 10^{-2}$    &    1.09$\pm$0.11  &   11.26$\pm$0.21     &  0.152$\pm$0.002  &    12.22$\pm$0.02        &     11.094$\pm$0.002   &  13.36$\pm$0.72      \\ 
1.00$\times 10^{-2}$    &    0.99$\pm$0.05  &   11.86$\pm$0.10     &  0.195$\pm$0.006  &    12.62$\pm$0.02        &     11.598$\pm$0.003   &  10.52$\pm$0.44      \\
3.16$\times 10^{-3}$    &    1.09$\pm$0.11  &   11.69$\pm$5.62    &  0.251$\pm$0.021  &    13.06$\pm$0.03        &     12.094$\pm$0.011   &  9.25$\pm$0.70       \\
\hline
\end{tabular}
\end{table*}

\section{Substructure occupancy variation}\label{apend:NsatNsub}

We show in Sec.~\ref{sec:HODenv} that 
haloes in the most dense environments are more likely to host 
a higher amount of satellites than those in the least dense 
environments, and in Sec.~\ref{sec:HODTime} that late formed 
haloes contain a higher mean number of satellites than early 
formed haloes.
Given these results, it is interesting to explore our findings 
in the context of the subhalo occupancy variation. This is particularly 
relevant for SHAM techniques. 

The simplest form of SHAM connects subhaloes with galaxies 
using a monotonic relation between a subhalo property such as 
the maximum circular velocity, V$_{\rm max}$, or the infall mass,
and a galaxy property like the stellar mass or luminosity. We note
however that most SHAM models tend to include some additional 
scatter in this relation to create a model that reproduces the 
data well enough. 

To some extent it should be expected that SHAM models will contain 
some subhaloe occupancy variation signal
\citep[e.g.,][]{Zentner2014,Chaves-Montero2016}. 
First, it is well established that
haloes with higher concentration have a larger V$_{\rm max}$ than less
concentrated ones \citep{Wechsler2002,Wechsler2006}. Moreover, haloes
with higher concentration assemble earlier. This makes it more likely 
for SHAM models to host galaxies in early formed haloes.  
Furthermore, haloes with higher concentrations contain less 
subhaloes compared to less concentrated ones
\citep{Zentner2005,Mao2015}, as subhaloes in early formed haloes have
more time to merge and/or deplete through dynamical friction.

In this appendix, we investigate the occupancy variation in the
context of subhaloes and galaxies. For this, we use the
\eagle\ simulation with the halo population split by environment and
formation time (see respectively \S~\ref{sec:defEnv} and 
\ref{sec:defFormTime} for further details). To compare the halo
occupancy of subhaloes and satellites, we adopt a cumulative number
density cut of n~=~0.0316~$h^3$~Mpc$^{-3}$, one of the thresholds 
used in \S~\ref{sec:results}. The two samples are created by ranking
satellite galaxies by stellar mass and subhaloes by V$_{\rm max}$. We 
note that the halos considered by both samples are not necessarily the 
same as there is some scatter in the subhalo V$_{\rm max}$ and
satellite stellar mass relation. 

\begin{figure*}
 \centering
   \includegraphics[width=0.9\textwidth]{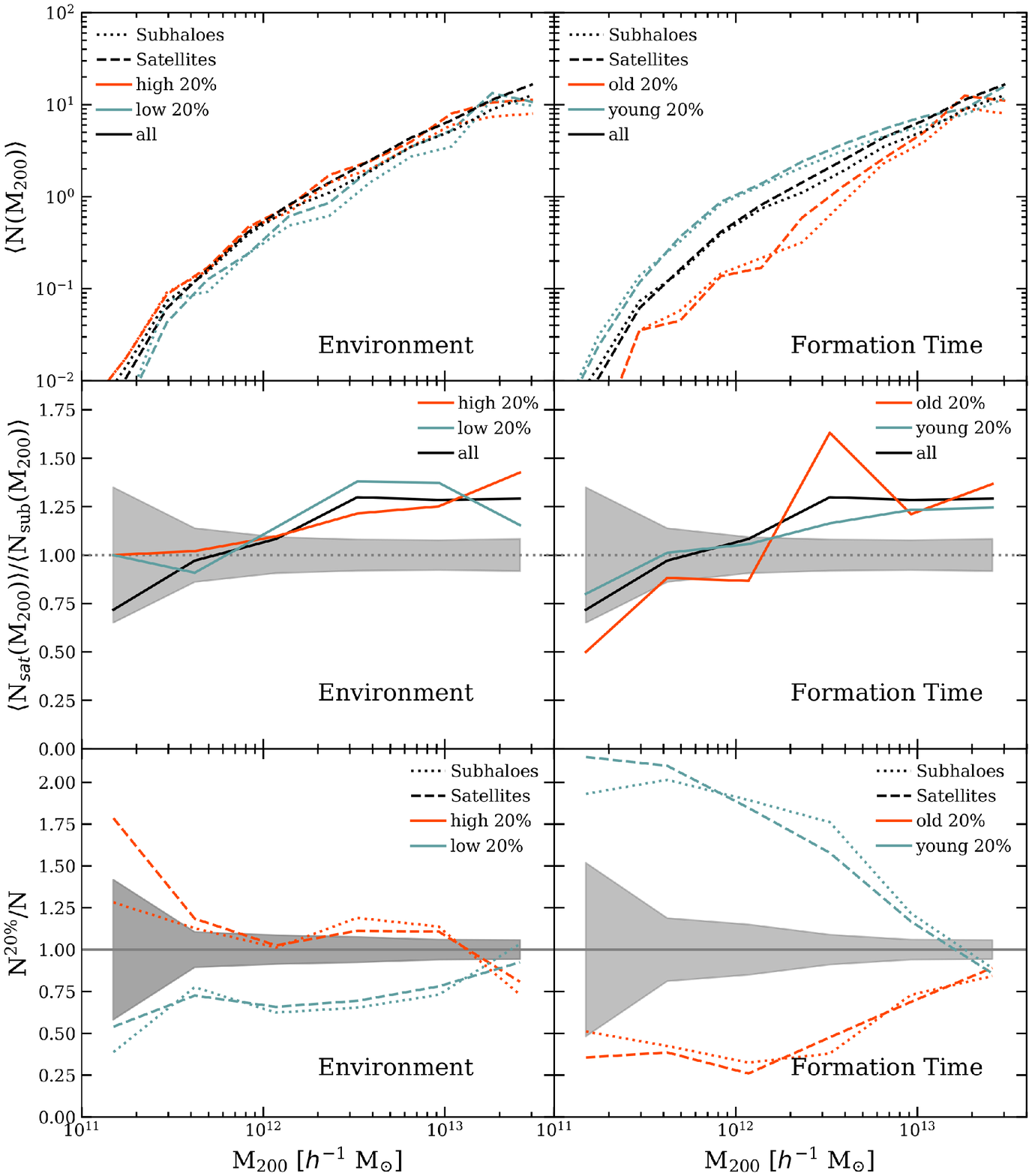}
  \caption{Comparison of satellite and subhalo occupancy in the
    \eagle\ simulation for a cumulative density cut of 
    n~$= 0.0316~h^3 {\rm Mpc}^{-3}$. The halo population is split by
    environment (left panel) and formation time (right panel; see
    \S~\ref{sec:defEnv} and \ref{sec:defFormTime} for their definitions).
    The top panels show the HOD for each sample of subhaloes (dotted
    lines) and satellites (dashed lines; see color code in the figure key). 
    The middle panels present the ratio between the HOD of satellites
    and subhaloes
    ($\langle$N$_{sat}$(M$_{200}$)$\rangle/\langle$N$_{\rm
      sub}$(M$_{200}$)$\rangle$) for the complete sample (black line)
    and according to environment and formation time (as shown by color
    code in panel key). The shaded grey area, centered on one, is used as
    reference to indicate the significance of the trends (see text for
    further details). 
    The bottom panels show the occupancy variation of satellites
    (dashed lines) and subhaloes (dotted lines) of the subsamples
    compared to the reference sample (N$^{20\%}$/N). The shaded grey
    area has the same role as in the middle panel (see text for
    further details). Our results show that the occupancy variations of 
    subhaloes in the \eagle\ simulation do not show any significant 
    differences with respect to the occupancy variations of satellite 
    galaxies.}
  \label{fig:NsatNsub}
 \end{figure*}

Fig.~\ref{fig:NsatNsub} shows our results split by halo environment
and halo formation time (left and right panels respectively). In the 
top row we present the HOD for subhaloes (dotted curves) and
satellites (dashed curves) of the \eagle\ simulation, including the
various sample splits (see figure key). Similar to our results for
satellites galaxies in Sec.~\ref{sec:HODenv} and
Sec.~\ref{sec:HODTime}, we find that haloes in denser environments (or
formed late) host on average more subhaloes than those in less dense
environments (or formed early). 
The middle row shows the ratio
of the average number of satellites per halo and the average number of
subhaloes per halo as function of halo mass (in effect the ratio of the
same coloured lines in the top panel). This ratio highlights further
differences in the HOD of satellites and subhaloes. To quantify their 
significance, we estimate the relative error on the ratio assuming
maximally anti-correlated Poisson 
statistics for the satellite and subhalo distributions. This provides
a simple upper limit to the relative error on the ratio, which for
convenience is plotted around the line of equal satellite and subhalo  
occupancy. At large halo masses 
(above $\sim 10^{12}~h^{-1}~{\rm~M}_{\odot}$), there are significantly more
satellites than subhaloes for the cumulative 
number density cut considered. This is primarily due to the existing
scatter in the subhalo V$_{\rm max}$ and satellite stellar mass
relation.

In the bottom panel of Fig.~\ref{fig:NsatNsub}, we show the
satellite/subhalo occupancy variation, as given by the ratio of the
HOD of satellites/subhaloes from subsamples split by formation time
(environment) and the HOD of satellites/subhaloes from the full
sample. Like in the middle panel, we estimate the significance
assuming Poisson statistics, but this time we consider the maximally
correlated Poisson statistic and plot the smallest one of the four
options as reference around the line of no occupancy variation. This 
provides a simple lower limit to the relative error on the ratios.
Hence in the \eagle\ simulation, we show that, for the cumulative
number density cut considered, the occupancy variations of subhaloes 
(dotted lines) do not show any significant differences with respect to
the occupancy variations of satellite galaxies (dashed lines).

\IfFileExists{\jobname.bbl}{}
{\typeout{}
\typeout{****************************************************}
\typeout{****************************************************}
\typeout{** Please run "bibtex \jobname" to optain}
\typeout{** the bibliography and then re-run LaTeX}
\typeout{** twice to fix the references!}
\typeout{****************************************************}
\typeout{****************************************************}
\typeout{}
}
\end{document}